\documentclass[english,twoside]{article}
\usepackage{slo,epsfig}
\usepackage{amsmath}
\usepackage{txfonts}
\usepackage{graphicx}
\usepackage{epstopdf}
\usepackage{multirow}
\usepackage{cancel}
\usepackage{comment}
\usepackage{tikz}
\usepackage{babel}

\newtheorem{prop}[theorem]{Proposition}

\newtheorem{defin}[theorem]{\protect\definname}

\newtheorem{remark}[theorem]{Remark}

\def\fl#1{\left\lfloor #1 \right\rfloor}
\def\ceil#1{\left\lceil #1 \right\rceil}

\def \drawOneU #1#2#3
{
\draw (#1, #2) -- (0.5+#1,#2) -- (0.5+#1,0.5+ #2) -- (1.5+#1,0.5+ #2)-- (1.5+#1,-0.5+ #2) -- (0.5+#1,-0.5+ #2) -- (0.5+#1,#2);
\draw (1.5+#1,#2) -- (2+#1,#2);
\draw (1+#1,#2) node{#3};
}

\def \drawTwoU #1#2#3#4
{
\draw (#1, #2) -- (0.3+ #1, #2);
\draw (#1, #3) -- (0.3+ #1, #3);
\draw (1.7+#1, #2) -- (2+ #1, #2);
\draw (1.7+#1, #3) -- (2+ #1, #3);

\draw (0.3+#1, -0.2+#2) -- (0.3+#1, 0.2+#3) -- (1.7+#1, 0.2+#3)-- (1.7+#1, -0.2+#2)-- (0.3+#1, -0.2+#2);

\draw (1+#1, 0.5* #2 + 0.5* #3) node{#4};
}

\def \drawThreeU #1#2#3#4#5
{
\draw (#1, #2) -- (0.5+ #1, #2);
\draw (#1, #3) -- (0.5+ #1, #3);
\draw (#1, #4) -- (0.5+ #1, #4);
\draw (1.5+#1, #2) -- (2+ #1, #2);
\draw (1.5+#1, #3) -- (2+ #1, #3);
\draw (1.5+#1, #4) -- (2+ #1, #4);

\draw (0.5+#1, -0.2+#2) -- (0.5+#1, 0.2+#4) -- (1.5+#1, 0.2+#4)-- (1.5+#1, -0.2+#2)-- (0.5+#1, -0.2+#2);

\draw (1+#1, 0.5* #2 + 0.5* #4) node{#5};
}

\def \drawSum #1#2#3
{
\draw (#1, #2) -- (2 + #1, #2);
\draw (#1, #3) -- (0.9+ #1, #3);
\draw (1.1+ #1, #3) -- (2+ #1, #3);

\draw (1 +#1, -0.5+ #2) -- (1+ #1, -0.1+#3);

\draw (1 +#1, #3) circle(0.1);

\draw (1 +#1, #2) circle(0.5);
}

\def \drawSumInv #1#2#3
{
\draw (#1, #2) -- (2 + #1, #2);
\draw (#1, #3) -- (0.9+ #1, #3);
\draw (1.1+ #1, #3) -- (2+ #1, #3);

\draw (1 +#1, -0.5+ #2) -- (1+ #1, -0.1+#3);

\draw (1 +#1, #3) circle(0.1);

\draw (1 +#1, #2) circle(0.5);

\draw (1.5 +#1, 0.5+ #2) node{\tiny{$\dag$}};
}

\def \drawCX #1#2#3
{
\draw (#1, #2) -- (2 + #1, #2);
\draw (#1, #3) -- (0.9+ #1, #3);
\draw (1.1+ #1, #3) -- (2+ #1, #3);

\draw (1 +#1, -0.5+ #2) -- (1+ #1, -0.1+#3);

\filldraw (1 +#1, #3) circle(0.1);

\draw (1 +#1, #2) circle(0.5);
}

\def \drawCXInv #1#2#3
{
\draw (#1, #2) -- (2 + #1, #2);
\draw (#1, #3) -- (0.9+ #1, #3);
\draw (1.1+ #1, #3) -- (2+ #1, #3);

\draw (1 +#1, -0.5+ #2) -- (1+ #1, -0.1+#3);

\filldraw (1 +#1, #3) circle(0.1);

\draw (1 +#1, #2) circle(0.5);

\draw (1.5 +#1, 0.5+ #2) node{\tiny{$\dag$}};
}

\def \drawSumPrime #1#2#3
{
\draw (#1, #3) -- (2 + #1, #3);
\draw (#1, #2) -- (0.9+ #1, #2);
\draw (1.1+ #1, #2) -- (2+ #1, #2);

\draw (1 +#1, 0.1+ #2) -- (1+ #1, 0.5+#3);

\draw (1 +#1, #2) circle(0.1);

\draw (1 +#1, #3) circle(0.5);
}

\def \drawSumPrimeInv #1#2#3
{
\draw (#1, #3) -- (2 + #1, #3);
\draw (#1, #2) -- (0.9+ #1, #2);
\draw (1.1+ #1, #2) -- (2+ #1, #2);

\draw (1 +#1, 0.1+ #2) -- (1+ #1, 0.5+#3);

\draw (1 +#1, #2) circle(0.1);

\draw (1 +#1, #3) circle(0.5);

\draw (1.5 +#1, 0.5+ #3) node{\tiny{$\dag$}};
}

\def \drawCXPrime #1#2#3
{
\draw (#1, #3) -- (2 + #1, #3);
\draw (#1, #2) -- (0.9+ #1, #2);
\draw (1.1+ #1, #2) -- (2+ #1, #2);

\draw (1 +#1, 0.1+ #2) -- (1+ #1, 0.5+#3);

\filldraw (1 +#1, #2) circle(0.1);

\draw (1 +#1, #3) circle(0.5);
}

\def \drawCXPrimeInv #1#2#3
{
\draw (#1, #3) -- (2 + #1, #3);
\draw (#1, #2) -- (0.9+ #1, #2);
\draw (1.1+ #1, #2) -- (2+ #1, #2);

\draw (1 +#1, 0.1+ #2) -- (1+ #1, 0.5+#3);

\filldraw (1 +#1, #2) circle(0.1);

\draw (1 +#1, #3) circle(0.5);

\draw (1.5 +#1, 0.5+ #3) node{\tiny{$\dag$}};
}

\def \Merge#1#2#3#4 
{
\draw (0.25+#1,#2) -- (0.25+#1,#4);
\draw [fill = white] (#1,-0.2+#4) rectangle (0.5+#1,0.2+#4);
\draw [fill = white] (#1,-0.2+#3) rectangle (0.5+#1,0.2+#3);
\draw [fill = white] (#1,-0.2+#2) -- (#1,0.2+#2) -- (0.5+#1,#2) -- (#1,-0.2+#2);

}

\def \MergeInv#1#2#3#4 
{
\draw (0.25+#1,#2) -- (0.25+#1,#4);
\draw [fill = white] (#1,-0.2+#4) rectangle (0.5+#1,0.2+#4);
\draw [fill = white] (#1,-0.2+#3) rectangle (0.5+#1,0.2+#3);
\draw [fill = white] (0.5+#1,-0.2+#2) -- (0.5+#1,0.2+#2) -- (#1,#2) -- (0.5+#1,-0.2+#2);

}

\def \CarryZero#1#2#3#4 
{
\draw (0.25+#1,#2) -- (0.25+#1,#4);
\draw [fill = black] (#1,-0.2+#4) rectangle (0.5+#1,0.2+#4);
\draw [fill = black] (#1,-0.2+#3) rectangle (0.5+#1,0.2+#3);
\draw [fill = white] (#1,-0.2+#2) -- (#1,0.2+#2) -- (0.5+#1,#2) -- (#1,-0.2+#2);

}

\def \CarryZeroInv#1#2#3#4 
{
\draw (0.25+#1,#2) -- (0.25+#1,#4);
\draw [fill = black] (#1,-0.2+#4) rectangle (0.5+#1,0.2+#4);
\draw [fill = black] (#1,-0.2+#3) rectangle (0.5+#1,0.2+#3);
\draw [fill = white] (0.5+#1,-0.2+#2) -- (0.5+#1,0.2+#2) -- (#1,#2) -- (0.5+#1,-0.2+#2);

}

\def \Carry#1#2#3 
{
\draw (0.25+#1,#2) -- (0.25+#1,#3);
\draw [fill = black] (#1,-0.2+#3) rectangle (0.5+#1,0.2+#3);
\draw [fill = white] (#1,-0.2+#2) -- (#1,0.2+#2) -- (0.5+#1,#2) -- (#1,-0.2+#2);

}

\def \CarryInv#1#2#3 
{
\draw (0.25+#1,#2) -- (0.25+#1,#3);
\draw [fill = black] (#1,-0.2+#3) rectangle (0.5+#1,0.2+#3);
\draw [fill = white] (0.5+#1,-0.2+#2) -- (0.5+#1,0.2+#2) -- (#1,#2) -- (0.5+#1,-0.2+#2);

}

\def \SmallSum#1#2#3 
{
\draw (0.25+#1,#3) -- (0.25+#1, -0.25+#2);
\draw (#1,#2) -- (0.5+#1, #2);
\draw (0.25+#1,#2) circle(0.25);
\draw [fill = white] (0.25+#1, #3) circle(0.1);
}

\def \SmallLambdaINCPrime#1#2#3 
{
\draw (0.25+#1,#2) -- (0.25+#1, 0.25+#3);
\draw (#1,#3) -- (0.5+#1, #3);
\draw (0.25+#1,#3) circle(0.25);
\draw [fill = white] (0.25+#1, #2) circle(0.1);
}

\def \DashLine#1#2#3 
{
\draw [dashed] (#1,#2) -- (#1+#3,#2);
}

\def \ArrowNote#1#2#3
{
\draw[->] (#1,#2) -- (#1, 0.8+#2);
\draw (#1, -0.5+#2) node{#3};
}

\providecommand{\definname}{Definition}
\providecommand{\observname}{Observation}
\providecommand{\corolname}{Corollary}
\providecommand{\examplename}{Example}

\def\C{{\mathbb{C}}}
\def\Z{{\mathbb{Z}}}

\def\F{{\mathbb{F}}}

\newcommand{\ket}[1]{|#1\rangle}

\newcommand{\nix}[1]{{}}

\newcommand{\CC}{\mathbb C}

\renewcommand{\phi}{\varphi}

\newcommand{\SU}{\rm SU}
\newcommand{\Horner}{\textrm{Horner}}

\def\CX{C(X)}
\def\CU#1{C(#1)}
\def\SoftU#1{\bigwedge(#1)}
\def\HardU#1#2{C_{#1}(#2)}
\def\SUM{{\sf{SUM}}}
\def\TwoSwap{S_{00,22}}
\def\CSUM{{C(\sf{SUM})}}

\def\ket#1{|#1\rangle}
\def\C{\mathcal{C}}

\def\M{\mathcal{M}}
\def\super{\text{supermetaplectic basis}}
\def\D{\mathcal{D}}
\def\SWAP{{\sf SWAP}}
\def\AdjC{{\sf AdjC}}
\def\F{\mathbb{F}}
\makeatother

\def\diag{\mathrm{diag}}
\def\Equ{\mathrm{Equ}}
\def\GL{\mathrm{GL}}
\def\square{\hbox{${\vcenter{\vbox{
   \hrule height 0.4pt\hbox{\vrule width 0.4pt height 6pt
   \kern5pt\vrule width 0.4pt}\hrule height 0.4pt}}}$}}

\begin{document}
\setlength{\textheight}{8.0truein}    

\runninghead{Improved Quantum Ternary Arithmetics}
            {A. Bocharov, S.X. Cui, M.Roetteler, K.M. Svore}

\normalsize\textlineskip
\thispagestyle{empty}
\setcounter{page}{1}

\vspace*{0.88truein}

\alphfootnote

\fpage{1}

\centerline{\bf
Improved Quantum Ternary Arithmetics}
\vspace*{0.37truein}

\centerline{\footnotesize
Alex Bocharov\footnote{alexeib@microsoft.com}}
\vspace*{0.015truein}
\centerline{\footnotesize\it Quantum Architectures and Computations Group, Microsoft Research}
\baselineskip=10pt
\centerline{\footnotesize\it Redmond, Washington, 98052, USA}
\vspace*{10pt}
\centerline{\footnotesize
Shawn X. Cui\footnote{cuixsh@gmail.com}}
\vspace*{0.015truein}
\centerline{\footnotesize\it University of California}
\baselineskip=10pt
\centerline{\footnotesize\it Santa Barbara, California, 93106,
USA}
\vspace*{10pt}
\centerline{\footnotesize
Martin Roetteler\footnote{martinro@microsoft.com}}
\vspace*{0.015truein}
\centerline{\footnotesize\it Quantum Architectures and Computations Group, Microsoft Research}
\baselineskip=10pt
\centerline{\footnotesize\it  Redmond, Washington, 98052, USA}
\vspace*{10pt}
\centerline{\footnotesize
Krysta M. Svore\footnote{ksvore@microsoft.com}}
\vspace*{0.015truein}
\centerline{\footnotesize\it Quantum Architectures and Computations Group, Microsoft Research}
\baselineskip=10pt
\centerline{\footnotesize\it Redmond, Washington, 98052, USA}
\vspace*{0.225truein}

\abstracts{
Qutrit (or ternary) structures arise naturally in many quantum systems, notably in certain non-abelian anyon systems.
We present efficient circuits for ternary reversible and quantum  arithmetics. Our main result is the derivation of circuits for two families of ternary quantum adders.
The main distinction from the binary adders is a richer ternary carry which leads potentially to higher resource counts in universal ternary bases.
Our ternary ripple adder circuit has a circuit depth of $O(n)$ and uses only $1$ ancilla, making it more efficient in both, circuit depth and width, when compared with previous constructions.
Our ternary carry lookahead circuit has a circuit depth of only $O(\log\,n)$, while using $O(n)$ ancillas.
Our approach works on two levels of abstraction: at the first level, descriptions of arithmetic circuits are given in terms of gates sequences that use various types of non-Clifford reflections. At the second level, we break down these reflections further by deriving them either from the two-qutrit Clifford gates and the non-Clifford gate $C(X): \ket{i,j}\mapsto \ket{i, j + \delta_{i,2} \; {\rm mod} \; 3}$ or from the two-qutrit Clifford gates and the non-Clifford gate $P_9=\mbox{diag}(e^{-2 \pi \, i/9},1,e^{2 \pi \, i/9})$. The two choices of elementary gate sets correspond to two possible mappings onto two different prospective quantum computing architectures which we call the metaplectic and the supermetaplectic basis, respectively. Finally, we develop a method to factor diagonal unitaries using multi-variate polynomials over the ternary finite field which allows to characterize classes of gates that can be implemented exactly over the supermetaplectic basis.
}{}{}

\vspace*{10pt}


\vspace*{1pt}\textlineskip    
\section{Introduction}
\label{sec:intro}
Quantum computation has seen vast progress over the years, both theoretically and experimentally. Computations on a programmable and scalable fault-tolerant quantum computer  will consist of fully controlled sequences of primitive operations such as unitary gates, measurements and state preparations. Such sequences are called \emph{quantum circuits}. In the most commonly used circuit model, quantum information is stored in a collection of \emph{qubits}, where each qubit has a two-dimensional Hilbert state space with the computational basis $\{\ket{0}, \ket{1}\}$. A standard universal gate set consists of Clifford gates and one non-Clifford gate such as the $\frac{\pi}{8}$-gate \cite{boykin2000new} or $V$-gate \cite{harrow2002efficient}. By design, circuits over a universal set can be used to approximate arbitrary quantum gates. Thus any quantum algorithm can be processed given a quantum computer with a universal gate set.

It has been noted by several researchers that architecture of certain quantum registers and gates is more naturally described by multi-valued logic as opposed to binary logic. History of experiments with ternary superconducting registers, in particular goes back to 1989 \cite{morisue1989jctl},\cite{morisue1998memory}. More recently, in quantum computation domain, multi-valued logic has been proposed for linear ion traps \cite{Muthu2000mvl}, cold atoms \cite{smith2013cs}, entangled photons \cite{malik2016multiPhoton}. It remains to be seen, at what scale it would be possible to balance out  quantum universality and fault-tolerance in these and other similar architectures.

The research presented here is motivated in part by recent progress in circuit synthesis over universal quantum bases arising in topological quantum computing, where multi-qubit encoding is not necessarily the most natural choice. Several physical systems capable of performing topologically-protected quantum computations have a natural structure of a qutrit instead of a qubit, where a qutrit has a three-dimensional Hilbert space with the computational basis $\{\ket{0},\ket{1},\ket{2}\}$. For instance, in the $\SU(2)_4$ anyon system, anyons with quantum dimension $\sqrt{3}$ are well-suited for encoding quantum states in qutrits. What is more, it was shown in \cite{cui2015universal} that the $\SU(2)_4$ anyon system can be made universal through braiding and projective measurement of anyons. This anyonic structure is quite far from physical realization at the moment, yet, it offers a promise of comparatively simple quantum universality combined with native topological protection, which, in our opinion, makes it a worthwhile subject of forward-looking research.


In \cite{bocharov2015efficient}, an algorithm is given for approximation of any multi-qutrit gate with an asymptotically optimal circuit over the gate set Clifford $+$ $\diag(1,1,-1)$. This work also demonstrated the importance of \emph{Householder reflections} for synthesis of efficient circuits. Even though the gate set turned out to be powerful enough for such synthesis, it had certain conceptual and practical limitations. Thus, it is quite  unlikely that all the reversible classical permutation gates can be implemented exactly over Clifford $+$ $\diag(1,1,-1)$. This has a damping effect on implementation of arithmetic-heavy algorithms such as Shor's Factorization Algorithm, since the integer modular arithmetic is naturally described by reversible classical circuits. As a matter of principle such circuits may be represented exactly in commonly used multi-qubit circuit models. \footnote{To the extent the three-qubit Toffoli gate may be assumed exactly representable.}


When compared to \cite{bocharov2015efficient}, the present paper aims at a more abstract level. Here we assume that the entire group of multi-qutrit classical permutations is representable at some cost, explore different scenarios of its representation and focus on synthesizing efficient circuits for ternary base arithmetic in these scenarios. Our thinking at this level remains reflection-centric.
Previous research on non-binary reversible circuits \cite{Brennen2006qudits}  mostly focused on proving the universality of the local classical Clifford gates in combination with the \emph{controlled-increment} gate $|j,k\rangle \mapsto |j, k + \delta_{j,d-1} \, {\rm mod} \; d\rangle$, where $d$ is the dimension of the qudit and $\delta$ is the Kronecker delta. Reversible circuits available in literature tend to use ancillary qudits fairly liberally.

This paper differentiates itself from previous work in two ways. First, we explore several alternate methods for synthesizing classical reversible circuits. Second, we strive to minimize both the depth and the width of arithmetic circuits specifically. For example, we show in Section \ref{subsec:ripple} that implementing of a faithful $\mbox{CARRY}$ gate is not necessary in a correct ternary adder. By using a modified carry we eliminate the use of ancillary qutrits and reduce the cost of the gate when compared to a faithful $\mbox{CARRY}$ as used in previous approaches to implement ternary carry ripple adders \cite{MDM:2004,satoh2007,khan2007quantum}.

Our focus is mainly on two types of ternary quantum adders, a modified ripple-carry adder and a carry look-ahead adder. Both adders are generalized from their binary counterparts, but the generalizations are somewhat non-trivial. To add two $n$-qutrit numbers, the modified ripple-carry adder uses $1$ ancilla and has a circuit depth of $O(n)$, while the carry look-ahead adder requires $O(n)$ ancillas and has a circuit depth of $O(\log \,n)$. Each of the two adders has an overall circuit size of $O(n)$ elementary gates.
We also study various extensions of quantum adders including adder modulo $3^n$, comparison, and subtraction.

We show these arithmetic circuits can be realized exactly using classical Clifford gates and one
additional
gate $\CX$, the \emph{controlled-increment} gate, whose matrix is given in Equation \ref{equ:CX}.
$\CX$ is a two-qutrit non-Clifford gate and it is universal for reversible classical computation. This sets the ternary reversible circuits apart from their binary analogs, where at least one three-qubit gate, e.g., the Toffoli gate, is required for universality.

\begin{equation}\label{equ:CX}
\CX =
\begin{pmatrix}
1 & 0& 0& 0& 0& 0& 0& 0& 0\\
0 & 1& 0& 0& 0& 0& 0& 0& 0\\
0 & 0& 1& 0& 0& 0& 0& 0& 0\\
0 & 0& 0& 1& 0& 0& 0& 0& 0\\
0 & 0& 0& 0& 1& 0& 0& 0& 0\\
0 & 0& 0& 0& 0& 1& 0& 0& 0\\
0 & 0& 0& 0& 0& 0& 0& 0& 1\\
0 & 0& 0& 0& 0& 0& 1& 0& 0\\
0 & 0& 0& 0& 0& 0& 0& 1& 0\\
\end{pmatrix}
\end{equation}

We also introduce a qutrit universal gate set Clifford $+$ $\diag(e^{-\frac{2\pi i}{9}}, 1, e^{\frac{2\pi i}{9}})$, called the supermetaplectic basis, which resembles the single-qubit $\frac{\pi}{8}$-gate. Some techniques are developed to construct new quantum gates from old ones. As an application, it will be shown that all ternary arithmetic studied in this paper can be implemented exactly over the $\super$.

We note that the reflection-centric synthesis of our adder circuits is a ternary counterpart of Toffoli-centric binary adder circuits as developed, for example, in \cite{cuccaro2004new} and \cite{draper2006logarithmic}. This analogy is explained in more detail in corresponding sections throughout the paper. The exact representation of the $\CX$ gate in $\super$ parallels the exact representation of the three-qubit Toffoli in the Clifford $+$ $\frac{\pi}{8}$ basis. Quantitative comparison of the ternary and binary adders would be beyond the scope of this work. A major step towards comprehensive comparison of this kind was made in the upcoming paper \cite{TernaryShor} that demonstrates the advantages of emulating Shor's period funding function on ternary quantum computer and especially on the metaplectic topological quantum framework.

The paper is organized as follows. In Section \ref{sec:background}, some preliminaries and notations used throughout the paper are given. In Section \ref{sec:adder}, we separately discuss the modified ripple-carry adder and carry look-ahead adder. Section \ref{sec:extensions} gives some extensions of quantum adders, including addition modulo $3^n$, comparison, and subtraction. Lastly in Section \ref{sec:techniques}, we introduce the $\super$ and develop techniques for the construction of new gates.

\section{Preliminaries and Notations}

\label{sec:background}
We denote the standard computational basis in a qutrit by $\{\ket{0}, \ket{1}, \ket{2}\}$. The terminology \lq\lq qutrit'' and \lq\lq ternary'' are sometimes used interchangeably. We call a quantum gate reversible or a classical \emph{permutation gate} if it acts as some permutation of the standard basis elements. Unless otherwise noted, the arithmetic, e.g., addition, multiplication, etc., within a ket is assumed to be taken modulo $3$. Also by default circuits are read from left to right, while compositions of gates when written as expressions follow the rule of matrix multiplications, i.e., they are read from right to left. Throughout the paper, the following ternary quantum gates are frequently used:
\begin{enumerate}
\item $X = \begin{pmatrix}
0 & 0 & 1\\
1 & 0 & 0 \\
0 & 1 & 0 \\
\end{pmatrix}$, namely, $X \ket{i} = \ket{i+1}$.

\item $S_{0,1} = \begin{pmatrix}
0 & 1 & 0\\
1 & 0 & 0 \\
0 & 0 & 1 \\
\end{pmatrix}$, namely, $S_{0,1}$ swaps $\ket{0}$ with $\ket{1}$ and fixes $\ket{2}$. Similarly, one can define $S_{0,2}, S_{1,2}$. This notation is also generalized to multi-qutrit gates. For instance, $\TwoSwap$ is a $2$-qutrit gate, which swaps $\ket{00}$ with $\ket{22}$, and fixes all other basis elements.

\item Given an $n$-qutrit gate $U$, there are two versions of \lq\lq controlled-$U$''. The first version is called \lq\lq soft-controlled-$U$,'' denoted by $\SoftU{U}$, and is defined as the $(n+1)$-qutrit gate: $\ket{i,j_1, \cdots j_n} \mapsto (I \otimes U^i)\ket{i,j_1, \cdots j_n}$, where the first qutrit is called the control qutrit. The second version is the \lq\lq hard-controlled-U'' denoted by $\HardU{c}{U}$, where $c \in \{0,1,2\}$. The gate $\HardU{c}{U}$ is also an $(n+1)$-qutrit gate. However, in contrast to the soft-controlled-$U$, it maps $\ket{i,j_1, \cdots j_n}$ to $(I \otimes U^{\delta_{i,c}})\ket{i,j_1, \cdots j_n}$. It is direct to see that the $\HardU{c}{U}\,'$s for different $c\,'$s are equivalent to each other up to some $1$-qutrit reversible gates. Thus we also use $\CU{U}$ to denote a general $\HardU{c}{U}$. Moreover, the equality $\SoftU{U} = \HardU{1}{U}(\HardU{2}{U})^2$ holds.

\item The following is a list of some important controlled gates:
      \begin{itemize}
        \item $\SUM = \SoftU{X}: \ket{i,j} \mapsto \ket{i,i+j}$,
        \item $\CX = \HardU{c}{X}: \ket{i,j} \mapsto \ket{i, j+\delta_{i,c}}$,
        \item $\Horner = \SoftU{\SoftU{X}}: \ket{i,j,k} \mapsto \ket{i,j,ij+k}$,
        \item $\CSUM = \HardU{c}{\SUM}: \ket{i,j,k} \mapsto \ket{i,j,j\delta_{i,c}+k}$.
      \end{itemize}
The $\Horner$ gate is a qutrit generalization of the qubit Toffoli gate. See also \cite{GRB:2003} for additional background on the Horner gate.
 \item $\SWAP: \ket{i,j} \mapsto \ket{j,i}$.
\end{enumerate}
For graphical representations of the gates defined above, see Figure \ref{fig:graph}.
\begin{figure}
\centering
\begin{tikzpicture}[scale = 0.5]

\begin{scope}
\draw (1,-1.5) node{$\SoftU{U}$};
\drawOneU{0}{0}{$U$};
\draw (0,2)--(2,2);
\draw (1,2) -- (1,0.5);
\filldraw[color = white] (1,2) circle(0.1);
\draw (1,2) circle(0.1);
\end{scope}

\begin{scope}[xshift = 3cm]
\draw (1,-1.5) node{$\HardU{c}{U}$};
\drawOneU{0}{0}{$U$};
\draw (0,2)--(2,2);
\draw (1,2) -- (1,0.5);
\filldraw (1,2) circle(0.1);
\draw (0.7,1.7) node{\tiny{$c$}};
\end{scope}

\begin{scope} [xshift =6cm]
\draw (1,-1.5) node{$X$};
\draw (0,0) -- (2,0);
\draw (1,0) circle(0.5);
\draw (1,-0.5) -- (1,0.5);
\end{scope}

\begin{scope}[xshift = 9cm]
\drawSum{0}{0}{2};
\draw (1,-1.5) node{$\SUM$};
\end{scope}

\begin{scope}[xshift = 12cm]
\drawCX{0}{0}{2};
\draw (1,-1.5) node{$\CX$};
\draw (0.8,1.7) node{\tiny{$c$}};
\end{scope}

\begin{scope}[xshift = 15cm]
\drawCX{0}{0}{2};
\draw (1,-1.5) node{$\CSUM$};
\draw (0.8,1.7) node{\tiny{$c$}};
\draw (0,1) -- (2,1);
\filldraw[color = white] (1,1) circle(0.1);
\draw (1,1) circle(0.1);
\end{scope}

\end{tikzpicture}
\fcaption{Graphical representations of some ternary gates}\label{fig:graph}
\end{figure}

The qutrit Clifford group $\C$ \cite{gottesman1999fault} is generated by $\SUM, X, H,$ and $Q$, where $H$ and $Q$ are defined as follows:
 $$ H = \frac{1}{\sqrt{3}}
 \begin{pmatrix}
 1 & 1 & 1 \\
 1 & \zeta_3 & \zeta_3^2 \\
 1 & \zeta_3^2 & \zeta_3 \\
 \end{pmatrix},
 \qquad
 Q = \begin{pmatrix}
 1 & 0 & 0 \\
 0 & 1 & 0 \\
 0 & 0 & \zeta_3 \\
 \end{pmatrix},
 $$
 where we use the notation $\zeta_n = e^{\frac{2\pi i}{n}}$ for $n\geq 1$.

 It can be shown that, along with the $\SUM$, all the reversible $1$-qutrit gates and $\SWAP$ are also contained in $\C$. Moreover, $\SUM$ and all the $1$-qutrit reversible gates generate the subgroup of all reversible gates in $\C$. Some other Clifford gates are $Z$ and $\bigwedge(Z)$, where $Z = \diag(1,\zeta_3,\zeta_3^2)$, and $\bigwedge(Z) = (I \otimes H)\SUM(I \otimes H^{-1}): \ket{i,j} \mapsto \zeta_3^{ij}\ket{i,j}$. However, $\CX, \Horner, \CSUM$ and $\TwoSwap$ are non-Clifford gates.

 Consider two pairs of standard basis vectors $|j_1\rangle, \,|k_1\rangle$ and $|j_2\rangle, \,|k_2\rangle$. It would be useful to note that the two-way classical reflection $S_{|j_1\rangle, |k_1\rangle}$ that swaps the $|j_1\rangle, \,|k_1\rangle$ and fixes everything else can be reduced to the corresponding reflection $S_{|j_2\rangle, |k_2\rangle}$ by applications of $O(n)$  $\, \SUM$ and $\SWAP$ gates (that are Clifford gates: see \cite{bocharov2015efficient}, Lemma 16). In particular, the two-way swap $\TwoSwap$ is Clifford-equivalent to any other two-qutrit two-way swap.


We think of Clifford gates as being {\it cheap} in the quantum sense. General rationale for this assumption is that such gates can be simulated classically. (Additional motivation coming from topological computing: in the context of non-abelian anyons such as
$\SU(2)_4$  anyon system \cite{cui2015universal}, Clifford gates can be obtained by anyon braiding alone.)
Thus we define the complexity (resp. depth) of a circuit as the number (resp. depth) of non-Clifford gates.

The following two identities will be used, where $\omega(n)$ is the number of $1\,'$s in the binary expansion of $n$, and $\fl{x}$ means the maximal integer less than or equal to $x$:

\begin{equation}
 \sum\limits_{i=1}^{\infty} \fl{\frac{n}{2^i}} = n - \omega(n),
\end{equation}

\begin{equation}
 \sum\limits_{i=1}^{\fl{\log\,n}+1} \fl{\frac{n}{2^i}-\frac{1}{2}} = n - \fl{\log \, n}  - 1.
\end{equation}

See also \cite{cuccaro2004new} for similar identities.

\section{Quantum Ternary Adders}
\label{sec:adder}
Given two $n$-trit numbers $a = a_{n-1} \cdots a_1a_0$, $b = b_{n-1} \cdots b_1b_0$, an adder computes their sum $s = s_n s_{n-1} \cdots s_0 = a+b$. The elementary method of adding two $n$-trit numbers is illustrated in Figure \ref{table:adder}. Let $c_0 = 0$ be the initial carry trit and for $1 \leq i \leq n$, let $c_{i}$ be the carry trit arising from $a_{i-1},b_{i-1}, c_{i-1}$, namely, $c_i = 0$ if $a_{i-1}+ b_{i-1} + c_{i-1} \leq 2$ and $c_i = 1$ otherwise. For $0 \leq i \leq n-1$, $s_i = a_i + b_i + c_i \, {\rm mod}\; 3$ and $s_n = c_n$.
\begin{table}
\centering
 \begin{tabular}{ccccl}
       &  $a_{n-1}$ & $\cdots$  & $a_{1}$  &  $a_{0}$  \\
       &  $b_{n-1}$ & $\cdots$  & $b_{1}$  &  $b_{0}$  \\
 $c_n$ &  $c_{n-1}$ & $\cdots$  & $c_{1}$  &  $c_{0}=0$  \\
 \hline
 $s_n$ &  $s_{n-1}$ & $\cdots$  & $s_{1}$  &  $s_{0}$ \\
 \end{tabular}
 \fcaption{Addition of two $n$-trit numbers}\label{table:adder}
\end{table}

In Section \ref{subsec:ripple} and Section \ref{subsec:lookahead}, we present  two methods to implement reversible ternary quantum adder: a ripple-carry adder and a carry look-ahead adder. The two adders are generalized from their binary counterparts \cite{cuccaro2004new,draper2006logarithmic}, but the generalizations are somewhat nontrivial, as seen later. On one hand, the modified ripple-carry adder uses only $1$ ancilla for the whole process and has the circuit depth in $O(n)$. On the other hand, the carry look-ahead adder requires $O(n)$ ancillas with the advantage of having circuit depth in $O(\log\,n)$. We will also compare the two adders to other ternary adders known in literature and show that our adders are more efficient both space-wise and depth-wise.

To implement the adders, we utilize $\CX$, $\CSUM$, $\CU{S_{0,1}}$ and $\TwoSwap$ as the basic building units. As shown in Section \ref{subsec:construction2}, $\CSUM$, $\CU{S_{0,1}}$ and $\TwoSwap$ can all be constructed exactly from $\CX$ and Clifford operations. Therefore, the circuit of adders can be designed from Clifford operations and $\CX$ alone. The reason that we still treat $\CSUM$, $\CU{S_{0,1}}$ and $\TwoSwap$ as basic units is that it might be more efficient to synthesize them directly in some basis rather than breaking them up into $\CX\,'$s. An example is the metaplectic basis \cite{bocharov2015efficient}, where $\TwoSwap$ has an efficient approximation by a metaplectic circuit.

 \subsection{Modified Ripple-Carry Adder}
 \label{subsec:ripple}
 The binary quantum ripple-carry adder was considered in \cite{vedral1996quantum}, where $O(n)$ ancillas are required to add two $n$-qubit numbers. In \cite{cuccaro2004new}, the method was improved so that only $1$ ancilla is necessary. Here we give a ternary generalization of the improved ripple-carry adder.

Note that in contrast to the binary case, the ternary carry is more complicated: if the inputs to a binary full adder are denoted by $a, b, c \in \F_2$, then the outgoing carry is given by $c_{out} = ab + ac + bc$, where all operations are computed modulo $2$. In case of a ternary full adder with inputs $a,b,c \in \F_3$, the outgoing carry is given by $c_{out} = 2(1+a+b+c)(ab+ac+bc)+abc$, where all operations are computed modulo $3$. Though directly implementing this polynomial using the presented universal gates is possible, it leads to a relatively large number of elementary gates. A simple observation allows to reduce this cost significantly as it turns out that $c_{out}$ does not have to be implemented for all $27$ input triples but rather only $18$ of them. Indeed, it can be shown inductively that---provided there is no initial incoming carry---for ternary adders, every carry trit $c_i$ can only be either $0$ or $1$, but can never be $2$. This is indicated also in Figure \ref{table:carry} where the crossed out case indicates that this can never occur in an actual addition: the case $c_{i+1} = 2$ is possible only if $c_i = 2$, which inductively we assume cannot happen. With this definition, $c_{i+1}$ becomes a balanced function, i.e., there are the same number of inputs corresponding to each outcome $c_{i+1}$.

We sketch the idea of constructing the circuit to compute $c_{i+1}$ from $a_i,b_i$ and $c_i$ based on this observation. As illustrated in Figure \ref{table:carry}, $c_{i+1}$ equals $c_{i}$ for all but six inputs, the last three inputs in the column $c_{i+1} = 0$ and the last three in the column $c_{i+1} = 1$. For each of these six inputs, $c_{i+1} $ equals $1-c_i$. If the gate $\TwoSwap$ is applied to qutrits $a_i,b_i$, then the six inputs are turned into six new triples. See Figure \ref{table:transition} for the transition. Moreover, the new six triples are exactly equal to the set $\{(a,b,c) \in \{0,1,2\}^3: a + b = c, c \neq 2\}$. In light of these observations, a reversible circuit, called Carry, is constructed, which takes $c_i, a_i,b_i$ as input, and outputs $c_{i+1}$ in the last qutrit. See Figure \ref{fig:carry}, where $f$ and $g$ are some functions of $a_i,b_i,c_i$. The exact shape of $f$ and $g$ is not important since they will be reversed at the appropriate step of the adder.

  \begin{table}
  \setlength{\tabcolsep}{4pt}
  \centering
    \begin{tabular}{c|ccccccccc|ccccccccc|ccccccccc}
          &  \multicolumn{9}{c|}{$c_{i+1}=0$}   &  \multicolumn{9}{c|}{$c_{i+1}=1$} &  \multicolumn{9}{c}{\cancel{$c_{i+1}=2$}}\\
    \hline
    $a_i$ & 0 & 0 & 0 & 1 & 1 & 2 & 0 & 0 & 1   & 0 & 1 & 1 & 2 & 2 & 2 & 1 & 2 & 2 & 0 & 0 & 0 & 1 & 1 & 1 & 2 & 2 & 2 \\
    $b_i$ & 0 & 1 & 2 & 0 & 1 & 0 & 0 & 1 & 0   & 2 & 1 & 2 & 0 & 1 & 2 & 2 & 1 & 2 & 0 & 1 & 2 & 0 & 1 & 2 & 0 & 1 & 2\\
    $c_i$ & 0 & 0 & 0 & 0 & 0 & 0 & 1 & 1 & 1   & 1 & 1 & 1 & 1 & 1 & 1 & 0 & 0 & 0 & 2 & 2 & 2 & 2 & 2 & 2 & 2 & 2 & 2   \\
    \end{tabular}
    \fcaption{Ternary carry table} \label{table:carry}
  \end{table}

  \begin{table}
\begin{tikzpicture}[scale = 0.5]
\begin{scope}
   \tikzstyle{block} = [draw, rectangle, minimum width=3cm, minimum height=1cm, text centered, text width=4.5cm,];
   \tikzstyle{arrow} = [thick,->,>=stealth];
   \tikzstyle{dasharrow} = [dashed,thick,->,>=stealth];
   \tikzstyle{invisibleblock} = [draw = none,rectangle, minimum width=3cm, minimum height=1cm, text centered, text width=8cm,];
    \node [invisibleblock](left){
    \begin{tabular}{c|ccc|ccc}
         & \multicolumn{3}{c|}{$c_{i+1}=0$}   &  \multicolumn{3}{c}{$c_{i+1}=1$} \\
    \hline
   $a_i$ & 0 & 0 & 1                          &  1 & 2 & 2 \\
   $b_i$ & 0 & 1 & 0                          &  2 & 1 & 2 \\
   $c_i$ & 1 & 1 & 1                          &  0 & 0 & 0 \\
    \end{tabular}
    };
     \node [invisibleblock, xshift = 6cm](right){
    \begin{tabular}{c|ccc|ccc}
         & \multicolumn{3}{c|}{$c_{i+1}=0$}   &  \multicolumn{3}{c}{$c_{i+1}=1$} \\
    \hline
   $a_i$ & 2 & 0 & 1                          &  1 & 2 & 0 \\
   $b_i$ & 2 & 1 & 0                          &  2 & 1 & 0 \\
   $c_i$ & 1 & 1 & 1                          &  0 & 0 & 0 \\
    \end{tabular}
    };

   \node [invisibleblock, xshift = 3cm](middle){
   $\overset{\TwoSwap}{\Longrightarrow}$
   };

\end{scope}
\end{tikzpicture}
\fcaption{Transition of inputs due to $\TwoSwap$}\label{table:transition}
\end{table}

 \begin{figure}
 \centering
 \begin{tikzpicture}[scale = 0.5]
   \begin{scope}
   \draw (0,3) node{\tiny{$c_i$}};
   \draw (0,1) node{\tiny{$b_i$}};
   \draw (0,2) node{\tiny{$a_i$}};

   \drawTwoU{1}{1}{2}{\tiny{$\TwoSwap$}};
   \draw (1,3) -- (2.5,3);

   \drawSum{2.7}{1}{2};
   \draw (2.5,3) -- (4,3);

   \drawSumInv{4}{1}{3};
   \draw (4,2) -- (4.8,2);
   \draw (5.2,2)--(5.5,2);
   \end{scope}

   \begin{scope}[xshift = 5.5cm]
   \drawOneU{0}{3}{\tiny{$S_{0,1}$}};
   \draw (0,1) -- (2,1);
   \draw (0,2) --(0.8,2);
   \draw (1.2,2)--(2,2);
   \filldraw (1,1) circle(0.1);
   \draw (1,1) -- (1,2.5);
   \draw (0.8,1.2) node{\tiny{0}};
    \end{scope}
    \begin{scope}[xshift = 7.5cm]
     \drawTwoU{0}{2}{3}{\tiny{$\SWAP$}};
     \draw (0,1)--(2,1);
     \drawTwoU{2}{1}{2}{\tiny{$\SWAP$}};
     \draw (2,3)--(4,3);
   \end{scope}
   \begin{scope}[xshift = 13cm]
   \draw (0,1) node{\tiny{$c_{i+1}$}};
   \draw (0.5,2) node{\tiny{$g(a_i,b_i,c_i)$}};
   \draw (0.5,3) node{\tiny{$f(a_i,b_i,c_i)$}};
   \end{scope}
 \end{tikzpicture}
 \fcaption{the circuit Carry}\label{fig:carry}
\end{figure}

As illustrated in Figure \ref{fig:carry}, the circuit Carry is ancilla free, in contrast to the carry circuit considered in \cite{satoh2007} where $1$ ancilla is required for each round of carry. See Figure \ref{fig:comparison} for the comparison. The circuit utilizes one $\TwoSwap$, one $\CU{S_{0,1}}$, two $\SUM$, and two $\SWAP$ gates. The $\SUM$ and $\SWAP$ are both Clifford gates, so only $2$ non-Clifford gates are needed. The depth of Carry in terms of non-Clifford gates is also $2$. Moreover, unlike the binary ripple-carry circuit MAJ \cite{cuccaro2004new} where the two qubits other than $c_{i+1}$ end up with $a_i+b_i, c_i+b_i$, in our circuit the two qutrits other than $c_{i+1}$  have the final values $f(a_i,b_i,c_i)$ and $g(a_i,b_i,c_i)$. This is the reason we call our carry circuit {\it modified}. However, as will be seen below, the modified carry circuit works in the same way as the regular one.

\begin{figure}
\centering
 \begin{tikzpicture}[scale = 0.5]
   \begin{scope}
     \draw (0,3) node{\tiny{$c_i$}};
     \draw (0,1) node{\tiny{$b_i$}};
     \draw (0,2) node{\tiny{$a_i$}};
     \draw (1,1) -- (2,1);
     \draw (1,2) -- (2,2);
     \draw (1,3) -- (2,3);
     \draw (2,0.8) -- (2,3.2) -- (5,3.2) -- (5,0.8) -- (2,0.8);
     \draw (5,1) -- (6,1);
     \draw (5,2) -- (6,2);
     \draw (5,3) -- (6,3);
     \draw (7,1) node{\tiny{$c_{i+1}$}};
     \draw (7.5,2) node{\tiny{$g(a_i,b_i,c_i)$}};
     \draw (7.5,3) node{\tiny{$f(a_i,b_i,c_i)$}};
     \draw (3.5,2) node{\tiny{Carry}};
    \draw[dashed] (9,-1) -- (9,5);
   \end{scope}

   \begin{scope} [xshift = 11cm, yshift = 1cm]
     \draw (0,0) node{\tiny{$0$}};
     \draw (0,3) node{\tiny{$c_i$}};
     \draw (0,1) node{\tiny{$b_i$}};
     \draw (0,2) node{\tiny{$a_i$}};
     \draw (1,0) -- (2,0);
     \draw (1,1) -- (2,1);
     \draw (1,2) -- (2,2);
     \draw (1,3) -- (2,3);
     \draw (2,-0.2) -- (2,3.2) -- (5,3.2) -- (5,-0.2) -- (2,-0.2);
     \draw (5,0) -- (6,0);
     \draw (5,1) -- (6,1);
     \draw (5,2) -- (6,2);
     \draw (5,3) -- (6,3);
     \draw (7,0) node{\tiny{$c_{i+1}$}};
     \draw (7,3) node{\tiny{$c_i$}};
     \draw (7,1) node{\tiny{$b_i$}};
     \draw (7,2) node{\tiny{$a_i$}};
     \draw (3.5,1.5) node{\tiny{Carry}};
   \end{scope}
 \end{tikzpicture}
 \fcaption{(Left) ripple carry in the present paper; (Right) ripple carry studied in \cite{satoh2007} }\label{fig:comparison}
\end{figure}

Let $C: \ket{c_i,a_i,b_i} \rightarrow \ket{f(a_i,b_i,c_i),g(a_i,b_i,c_i), c_{i+1}}$ be the Carry gate represented by the circuit in Figure \ref{fig:carry}. Similar to the adder circuit in \cite{cuccaro2004new}, the modified ripple-carry adder circuit is designed in Figure \ref{fig:rippleadder}, which, as an illustration, shows the addition of two $3$-qutrit numbers.

\begin{figure}
\centering
 \begin{tikzpicture}[scale = 0.5]
   \begin{scope}
     \draw (0,0) node{$0$};
     \draw (0,1) node{$b_2$};
     \draw (0,2) node{$a_2$};
     \draw (0,3) node{$b_1$};
     \draw (0,4) node{$a_1$};
     \draw (0,5) node{$b_0$};
     \draw (0,6) node{$a_0$};
     \draw (0,7) node{$c_0$};
   \end{scope}

   \begin{scope}[xshift = 1cm]
     \drawThreeU{0}{5}{6}{7}{\tiny{$C$}};
     \draw (0,0) -- (1.5,0);
     \draw (0,1) -- (1.5,1);
     \draw (0,2) -- (1.5,2);
     \draw (0,3) -- (1.5,3);
     \draw (0,4) -- (1.5,4);
     \begin{scope}[xshift = 1.5cm]
     \drawThreeU{0}{3}{4}{5}{\tiny{$C$}};
     \draw (0,0) -- (1.5,0);
     \draw (0,1) -- (1.5,1);
     \draw (0,2) -- (1.5,2);
     \draw (0,6) -- (1.5,6);
     \draw (0,7) -- (1.5,7);
     \begin{scope}[xshift = 1.5cm]
     \drawThreeU{0}{1}{2}{3}{\tiny{$C$}};
     \draw (0,0) -- (1.5,0);
     \draw (0,4) -- (1.5,4);
     \draw (0,5) -- (1.5,5);
     \draw (0,6) -- (1.5,6);
     \draw (0,7) -- (1.5,7);
     \begin{scope}[xshift = 1.5cm]
     \drawSum{0}{0}{1};
     \draw (0,2) -- (1.5,2);
     \draw (0,3) -- (1.5,3);
     \draw (0,4) -- (1.5,4);
     \draw (0,5) -- (1.5,5);
     \draw (0,6) -- (1.5,6);
     \draw (0,7) -- (1.5,7);
     \begin{scope}[xshift = 1.5cm]
     \drawThreeU{0}{1}{2}{3}{\tiny{$C^{-1}$}};
     \draw (0,0) -- (1.5,0);
     \draw (0,4) -- (1.5,4);
     \draw (0,5) -- (1.5,5);
     \draw (0,6) -- (1.5,6);
     \draw (0,7) -- (1.5,7);
     \begin{scope}[xshift = 1.5cm]
     \drawSum{0}{1}{2};
     \draw (0,0) -- (1.5,0);
     \draw (0,3) -- (1.5,3);
     \draw (0,4) -- (1.5,4);
     \draw (0,5) -- (1.5,5);
     \draw (0,6) -- (1.5,6);
     \draw (0,7) -- (1.5,7);
     \begin{scope}[xshift = 1.5cm]
     \drawSum{0}{1}{3};
     \draw (0,0) -- (1.5,0);
     \draw (0,2) -- (0.8,2);
     \draw (1.2,2) -- (1.5,2);
     \draw (0,4) -- (1.5,4);
     \draw (0,5) -- (1.5,5);
     \draw (0,6) -- (1.5,6);
     \draw (0,7) -- (1.5,7);
     \begin{scope}[xshift = 1.5cm]
     \drawThreeU{0}{3}{4}{5}{\tiny{$C^{-1}$}};
     \draw (0,0) -- (1.5,0);
     \draw (0,1) -- (1.5,1);
     \draw (0,2) -- (1.5,2);
     \draw (0,6) -- (1.5,6);
     \draw (0,7) -- (1.5,7);
     \begin{scope}[xshift = 1.5cm]
     \drawSum{0}{3}{4};
     \draw (0,0) -- (1.5,0);
     \draw (0,1) -- (1.5,1);
     \draw (0,2) -- (1.5,2);
     \draw (0,5) -- (1.5,5);
     \draw (0,6) -- (1.5,6);
     \draw (0,7) -- (1.5,7);
     \begin{scope}[xshift = 1.5cm]
     \drawSum{0}{3}{5};
     \draw (0,0) -- (1.5,0);
     \draw (0,2) -- (0.8,2);
     \draw (1.2,2) -- (1.5,2);
     \draw (0,4) -- (0.8,4);
     \draw (1.2,4) -- (1.5,4);
     \draw (0,1) -- (1.5,1);
     \draw (0,2) -- (1.5,2);
     \draw (0,6) -- (1.5,6);
     \draw (0,7) -- (1.5,7);
     \begin{scope}[xshift = 1.5cm]
     \drawThreeU{0}{5}{6}{7}{\tiny{$C^{-1}$}};
     \draw (0,0) -- (1.5,0);
     \draw (0,1) -- (1.5,1);
     \draw (0,2) -- (1.5,2);
     \draw (0,3) -- (1.5,3);
     \draw (0,4) -- (1.5,4);
     \begin{scope}[xshift = 1.5cm]
     \drawSum{0}{5}{6};
     \draw (0,0) -- (2,0);
     \draw (0,1) -- (2,1);
     \draw (0,2) -- (2,2);
     \draw (0,3) -- (2,3);
     \draw (0,4) -- (2,4);
     \draw (0,7) -- (2,7);
     \begin{scope}[xshift = 2.5cm]
     \draw (0,0) node{$s_3$};
     \draw (0,1) node{$s_2$};
     \draw (0,2) node{$a_2$};
     \draw (0,3) node{$s_1$};
     \draw (0,4) node{$a_1$};
     \draw (0,5) node{$s_0$};
     \draw (0,6) node{$a_0$};
     \draw (0,7) node{$c_0$};
   \end{scope}
   \end{scope}
   \end{scope}
   \end{scope}
   \end{scope}
   \end{scope}
   \end{scope}
   \end{scope}
   \end{scope}
   \end{scope}
   \end{scope}
   \end{scope}
   \end{scope}

 \end{tikzpicture}
 \fcaption{Circuit for ripple-carry adder}\label{fig:rippleadder}
\end{figure}

In Figure \ref{fig:rippleadder}, the qutrit $c_0$, initialized with $0$, is the only ancilla required. The qutrit on the bottom holds the overflow trit, i.e., the highest trit in the sum. Therefore, to add two $n$-qutrit numbers, exactly $1$ ancilla, $n$ Carry gates, $n$ inverse Carry gates and $2\,n$ $\SUM$ gates are required, and the depth of the circuit is $4\,n$. In contrast, the adder in \cite{satoh2007} uses $n$ ancillas and has the complexity in $O(n)$.


 \subsection{Carry Look-ahead Adder}
 \label{subsec:lookahead}

In the ripple-carry adder, the carry $c_{i+1}$ is computed only after the value of $c_i$ has been obtained, and thus the overall depth of the circuit is in $O(n)$. One protocol to reduce the depth is the carry look-ahead adder studied in \cite{draper2006logarithmic} for the binary addition. Here we generalize it to give a ternary carry look-ahead adder, which computes all the carry trits in depth $O(\log\,n)$ by introducing extra $O(n)$ ancillas.

The main idea is that there are relations between $c_i$ and $c_{i+1}$, and more generally between $c_i$ and $c_j$ for $i \neq j$. For instance, if $a_i + b_i = 2$, then $c_{i+1}= c_i$. If $a_i + b_i = 1$, then $c_{i+1} = 0$ regardless of the value of $c_i$. See Figure \ref{table:relation} for a summary of the relation between $c_{i+1}$ and $c_{i}$. Note that $c_0 = 0$, thus when $i=0$, the column $c_{i+1}=c_i$ in Figure \ref{table:relation} becomes $c_1 = c_0 = 0$. Motivated by their relations, we define, for $0 \leq i < j \leq n$, the carry status indicator $C[i,j]:$

 \begin{table}
 \centering
 \begin{tabular}{c|ccc|ccc|ccc}
       & \multicolumn{3}{c|}{$c_{i+1} = 0$}  & \multicolumn{3}{c|}{$c_{i+1} = 1$}  & \multicolumn{3}{c}{$c_{i+1} = c_i$} \\
  \hline
 $a_i$ & 0 & 0 & 1                          & 1 & 2 & 2                          & 0 & 1 & 2                           \\
 $b_i$ & 0 & 1 & 0                          & 2 & 1 & 2                          & 2 & 1 & 0                           \\
 \end{tabular}
 \fcaption{Relation between $c_{i+1}$ and $c_i$}\label{table:relation}
\end{table}

$$C[i,j] = \begin{cases}
0         & c_{j} = 0 \text{ regardless of }c_i \\
1         & c_{j} = 1 \text{ regardless of }c_i \\
2         & c_{j} = c_i \\
\end{cases}$$

Since we already know $c_0 = 0$, the case $c_j = c_0$ is then the same as the first case $c_j = 0$. Thus we can treat these two cases as one, and design $C[0,j]$ so that it will never take the value $2$, namely, we will have $C[0,j] = c_j$.

Explicitly, for $0 < i < n$, the circuit, $\AdjC$, shown in Figure \ref{fig:Ci} computes $C[i,i+1]$ from $a_i$ and $b_i$. It requires $1$ non-Clifford gate $\TwoSwap$, and no ancilla. However, to compute $C[0,1]$, we need to make use of $1$ ancilla, and $2$ non-Clifford gates $\TwoSwap, \CX$. See Figure \ref{fig:C0} for the circuit, which we call $\AdjC_0$.

\begin{figure}
\centering
 \begin{tikzpicture}[scale = 0.5]
   \begin{scope}
     \draw (0,0) node {\tiny{$b_i$}};
     \draw (0,1) node {\tiny{$a_i$}};
     \begin{scope}[xshift = 1cm]
        \drawTwoU{0}{0}{1}{\tiny{$S_{00,22}$}};
        \begin{scope}[xshift = 2cm]
           \drawSum{0}{0}{1};
           \begin{scope}[xshift = 2cm]
             \drawOneU{0}{0}{\tiny{$S_{0,1}$}};
             \draw (0,1) -- (2,1);
             \begin{scope}[xshift = 2.5cm]
               \draw (1,0) node {\tiny{$C[i,i+1]$}};
               \draw (1,1) node { };
             \end{scope}
           \end{scope}
        \end{scope}
     \end{scope}
   \end{scope}
 \end{tikzpicture}
 \fcaption{Circuit $\AdjC$ computing $C[i,i+1]$, $0 < i < n$}\label{fig:Ci}
\end{figure}

\begin{figure}
\centering
 \begin{tikzpicture}[scale = 0.5]
   \begin{scope}
     \draw (0,0) node {\tiny{$b_i$}};
     \draw (0,1) node {\tiny{$a_i$}};
     \draw (0,2) node {\tiny{$0$}};
     \begin{scope}[xshift = 1cm]
        \drawTwoU{0}{0}{1}{\tiny{$S_{00,22}$}};
        \draw (0,2) -- (2,2);
        \begin{scope}[xshift = 2cm]
           \drawSum{0}{0}{1};
           \draw (0,2) -- (2,2);
           \begin{scope}[xshift = 2cm]
             \drawCXPrime{0}{0}{2};
             \draw (0.6,0.2) node {\tiny{$0$}};
             \draw (0,1) -- (0.8,1);
             \draw (1.2,1) -- (2,1);
             \begin{scope}[xshift = 2cm]
               \drawTwoU{0}{1}{2}{\tiny{$\SWAP$}};
               \draw (0,0) -- (2,0);
               \begin{scope}[xshift = 2cm]
               \drawTwoU{0}{0}{1}{\tiny{$\SWAP$}};
               \draw (0,2) -- (2,2);
               \begin{scope}[xshift = 2cm]
               \draw (1,2) node { };
               \draw (1,1) node { };
               \draw (1,0) node {\tiny{$C[0,1]$}};
               \end{scope}
               \end{scope}
             \end{scope}
           \end{scope}
        \end{scope}
     \end{scope}
   \end{scope}
 \end{tikzpicture}
 \fcaption{Circuit $\AdjC_0$ computing $C[0,1]$}\label{fig:C0}
\end{figure}



Having computed the carry status indicators for any two adjacent indices, we furthermore compute $C[i,j]$ for arbitrary $i \neq j$. For $0 \leq i < k < j \leq n$, $C[i,j]$ can be obtained from $C[i,k]$ and $C[k,j]$ by the {\it merging} formula in Figure \ref{table:merging}.

\begin{table}
\centering
 \begin{tabular}{cc|ccc}
   \multicolumn{2}{c|}{\multirow{2}{*}{$\bigodot$}}    &  \multicolumn{3}{c}{$C[k,j]$} \\
    \multicolumn{2}{c|}{\multirow{2}{*}{ }}                             &  0  & 1  &  2                 \\
  \hline
  \multirow{3}{*}{$C[i,k]$}  & 0  &  0  & 1  & 0                  \\
                             & 1  &  0  & 1  & 1                  \\
                             & 2  &  0  & 1  & 2                  \\
 \end{tabular}
 \fcaption{The {\it merging} formula $C[i,j] =  C[i,k] \bigodot C[k,j]$} \label{table:merging}
\end{table}

Note that when $i=0$, the row corresponding to $C[0,k] = 2$ in Figure \ref{table:merging} will never be used. Also when $C[0,k]$ takes values in $\{0,1\}$, so will $C[0,j]$. A circuit, $\M$, realizing the {\it merging} formula is illustrated in Figure \ref{fig:merging}, where $\M$ takes $C[i,k], C[k,j]$, and an ancilla initialized to $0$ as inputs, and outputs $C[i,j]$ to the ancilla. The circuit requires $1$ non-Clifford gate $\CSUM$.

\begin{figure}
\centering
 \begin{tikzpicture}[scale = 0.5]
   \begin{scope}
     \draw (0,0) node {\tiny{$0$}};
     \draw (0,1) node {\tiny{$C[k,j]$}};
     \draw (0,2) node {\tiny{$C[i,k]$}};
     \begin{scope}[xshift = 1cm]
       \drawSum{0}{0}{1};
       \draw (0,2) -- (2,2);
       \begin{scope}[xshift = 2cm]
          \drawSumPrimeInv{0}{1}{2};
          \draw (0,0) -- (2,0);
          \begin{scope}[xshift = 2cm]
             \drawSum{0}{0}{2};
             \draw (0,1) -- (2,1);
             \filldraw (1,1) circle(0.1);
             \draw (0.8,0.8) node {\tiny{$2$}};
             \begin{scope}[xshift = 2cm]
                \drawSumPrime{0}{1}{2};
                \draw (0,0) -- (2,0);
                \begin{scope}[xshift = 2cm]
                  \draw (1,0) node {\tiny{$C[i,j]$}};
                  \draw (1,1) node {\tiny{$C[k,j]$}};
                  \draw (1,2) node {\tiny{$C[i,k]$}};
                \end{scope}
             \end{scope}
          \end{scope}
       \end{scope}
     \end{scope}
   \end{scope}
 \end{tikzpicture}
 \fcaption{Circuit $\M$ realizing the {\it merging} formula}\label{fig:merging}
\end{figure}

The circuits $\AdjC$ and $AdjC_0$ both only depend on $a_i$ and $b_i$, thus we can compute all the $C[i,i+1]\,'$s in one time slice. The nature of the {\it merging} formula enables us to obtain all the $C[0,j]\,'$s in $O(\log \, n)$ time slices. We elaborate this below.

For $i = 0,1,\cdots,n-1$, let $B_i$ be the working register configured to be $C[i,i+1]$ at the beginning, and let $ Z_{i+1}$ be the working registers initialized to $\ket{0}$, which will end up with $C[0,i+1]$. We also need $n-\omega(n) - \fl{\log\,n}$ ancillas $X_i$ initialized to $\ket{0}$. The circuit consists of three processes, namely, $P$-process, $C$-process, and $P^{-1}$-process. Each process roughly contains $\fl{\log \,n}$ rounds.

In $P$-process, we compute all the carry status indicators of the form $C[2^t m, 2^t (m+1)]$ and write all the results into the ancillas, except the ones $C[0,2^k]$ which are written to $Z[2^k]$. There are $\fl{\log\,n}$ rounds, each $t = 1, \cdots, \fl{\log\,n}$ corresponding to one round. In the $t$-th round, which we call the $P[t]$-round, the status indicators $C[2^t m, 2^t(m+1)]$, $m = 0, \cdots, \fl{\frac{n}{2^t}}-1$ are computed. By the {\it merging} formula, $C[2^t m, 2^t (m+1)]$ can be obtained from $C[2^{t-1}(2m),2^{t-1}(2m+1)]$ and $[2^{t-1}(2m+1), 2^{t-1}(2m+2)]$, both of which have been computed in $P[t-1]$-round by induction. Moreover, the circuit $\M$ producing $C[2^t m, 2^t (m+1)]$ for different $m\,'$s in the $P[t]$-round takes different carry status indicators in $P[t-1]$-round as input. Note that the $P[1]$-round requires the carry status indicators $C[i,i+1]\,'$s in the registers $B_i$. Therefore, in the $P[t]$-round, all the circuits $\M$ computing $C[2^t m, 2^t (m+1)]$ can be made parallel, and their inputs only depend on the carry status indicators from the $P[t-1]$-round. Thus, the depth of the circuit in $P$-process is $\fl{\log\,n}$, the number of ancillas needed is $n-\omega(n) - \fl{\log\,n}$, and the complexity is $n-\omega(n)$.


In $C$-process, we compute $C[0,j]$ into the register $Z_j$, $j = 1, \cdots, n$. This is performed in $\fl{\log\,\frac{n}{3}}+1$ rounds. Note that the $C[0,2^k] \,'$s have already been obtained in $P$-process, and are located in the desired positions. For $t = \fl{\log\,\frac{n}{3}}, \cdots , 0$, the $C[t]$-round consists of computing the carry status indicators $C[0, 2^t(2m+1)],\, m = 1, \cdots, \fl{\frac{n}{2^{t+1}}-\frac{1}{2}}$. Again, by the {\it merging} formula, we can get $C[0,2^t(2m+1)]$ from $C[0, 2^{t+1}m]$ and $C[2^{t}(2m), 2^{t}(2m+1)]$. By induction, $C[0, 2^{t+1}m]$ has been obtained in earlier $C$-rounds if $m$ is not a power of $2$, and in the $P[t+1+ \log \,m]$-round otherwise. Also $C[2^{t}(2m), 2^{t}(2m+1)]$ has been computed in the $P[t]$-round. Therefore, we can run all the $\M$ circuits in the $C[t]$-round in a parallel way. These circuits depend on the carry status indicators in the $P[t]$-round and $C[k]$-rounds, $k \geq t+1$. If $m$ is a power of $2$, then the corresponding $\M$ circuit also depends on $C[0,2^{t+1}m]$ from the $P[t+1+\log\,m]$-round. Thus the circuit in $C$-process has a depth of $\fl{\log\,\frac{n}{3}}+1$, and the complexity is $n-\fl{\log\,n} -1$.


In $P^{-1}$-process, we set the ancillas back to $\ket{0}$, thus we need to reverse all the $\M$ circuits in $P$-process, except for those computing $C[0,2^k]\,'$s which are not stored in the ancillas. The $P^{-1}$-process consists of $\fl{\log\,n}-1$ rounds. For $t = \fl{\log\,n}-1, \cdots, 1$, the $P^{-1}[t]$-round uncomputes $C[2^t m, 2^t (m+1)]$, $m = 1, \cdots, \fl{\frac{n}{2^t}}-1$ by using the inverse of $\M$. Note that in this process, all the $\,C[0,2^k]'$s will not be touched. The process has a depth of $\fl{\log\,n}-1$, and the complexity of the circuit is $n-\omega(n) - \fl{\log\,n}$.


We note that most parts of $C$-process and $P^{-1}$-process can actually be parallelized. The argument is as follows. All the inputs to the $C[t]$-round which are not of the form $C[0,2^m]$ only depend on $C[k]$-rounds, $k\geq t+1$, and the $P[t]$-round. The inputs that are of the form $C[0,2^m]$ were computed in $P[m]$-round, but they will not be touched in $P^{-1}$-process. The $P^{-1}[t+2]$-round only depends on the outputs in $P[t+1]$-round and $P[t+2]$-round. Thus the $C[t]$-round and the $P^{-1}[t+2]$-round can be performed simultaneously. The precise parallelism between $C$-process and $P^{-1}$-process is illustrated in Figure \ref{table:parallel}.
\begin{table}
\centering
  \begin{tabular}{c|c|c|c|c|c}
    $C[\fl{\log\,\frac{n}{3}}]$  &  $\cdots$  &   $C[\fl{\log \,n}-3]$      & $\cdots$  & $C[0]$       &    \\
                                 &            &   $P^{-1}[\fl{\log \,n}-1]$ & $\cdots$  & $P^{-1}[2]$  & $P^{-1}[1]$ \\
  \end{tabular}
  \fcaption{Parallelism between $C$- and $P^{-1}$-process}\label{table:parallel}
\end{table}

To summarize, the whole circuit uses $n-\omega(n) - \fl{\log\,n}$ ancillas, and has a depth of $\fl{\log\,n} + \fl{\log\,\frac{n}{3}} +2$. The total complexity of the circuit is $3n-2\omega(n) - 2\fl{\log\,n}  -1 $.


 \subsection{Complete Circuit for Carry Look-Ahead Adder}
We give two implementations of carry look-ahead adder, namely, the out-of-place adder and the in-place adder. Recall that the circuits in Figure \ref{fig:Ci}, \ref{fig:C0}, and \ref{fig:merging} are denoted by $\AdjC$, $\AdjC_0$, and $\M$, respectively. The complexity of both $\AdjC$ and $\M$ is $1$, and the complexity of $\AdjC_0$ is $2$. The depth of these circuits is equal to their complexity.

 \subsubsection{Out-of-place Adder}
\label{subsubsec:out-of-place}

Let $A_i, B_i$ be the registers with initial value $a_i, b_i$, respectively, $i = 0, \cdots, n-1$. Let $Z_i, i = 0, \cdots, n$ be the registers initialized to be $0$, which will hold the sum $a+b$ at the end of the computation. We need $n-\omega(n) - \fl{\log\,n}$ ancillas $X_i$ to store intermediate carry status indicators. The following is a description of the circuit of our out-of-place adder.

\vspace{0.5cm}
\textbf{Out-of-place Procedure:}

\begin{enumerate}
 \item For $ 0 < i \leq n-1$, run the circuit $\AdjC$ on $A_i, B_i$, which outputs $C[i,i+1]$ to $B_i$. Run $\AdjC_0$ on $A_0,B_0$, and $Z_0$ with $Z_0$ as the ancilla, which outputs $C[0,1]$ to $B_0$. Copy $C[0,1]$ to $Z_1$ with the $\SUM$ gate. The circuit has a depth of $2$, and it consist of $n-1$ $\AdjC$, $1$ $\AdjC_0$, and $1$ $\SUM$ gates.
 \item As discussed in Section \ref{subsec:lookahead}, compute all the $C[0,i]\,'$s with the ancillas $X_i\,'$s and the circuit $\M^{\pm 1}$. At the end of this process, the ancillas are returned to $0$, and $Z_i = C[0,i], i = 1, \cdots, n$.\footnote{$Z_1 = C[0,1]$ was obtained in the previous step.} This requires $3n-2\omega(n) - 2\fl{\log\,n}  -1 $ calls to the circuit $\M^{\pm 1}$, and has a circuit depth of $\fl{\log\,n} + \fl{\log\,\frac{n}{3}} +2$.
 \item Undo all the $\AdjC\,'$s and $\AdjC_0$. At the end of this step, we have $B_i = b_i, Z_i = C[0,i] = c_i.$ The circuit has a depth of $2$, and it consist of $n-1$ $\AdjC^{-1}$, $1$ $\AdjC_0^{-1}$, and $1$ $\SUM^{-1}$.
 \item Set $Z_i = Z_i \oplus A_i \oplus B_i$, $0\leq i \leq n-1$. This requires $2n$ $\SUM$ gates.
\end{enumerate}
\vspace{0.5cm}

In summary, the out-of-place adder uses $n-\omega(n) - \fl{\log\,n}$ ancillas, and has a circuit depth of $\fl{\log\,n} + \fl{\log\,\frac{n}{3}} +6$, with the  complexity of $5n-2\omega(n) - 2\fl{\log\,n}  -1$.

\begin{figure}
\includegraphics[width = 1\textwidth]{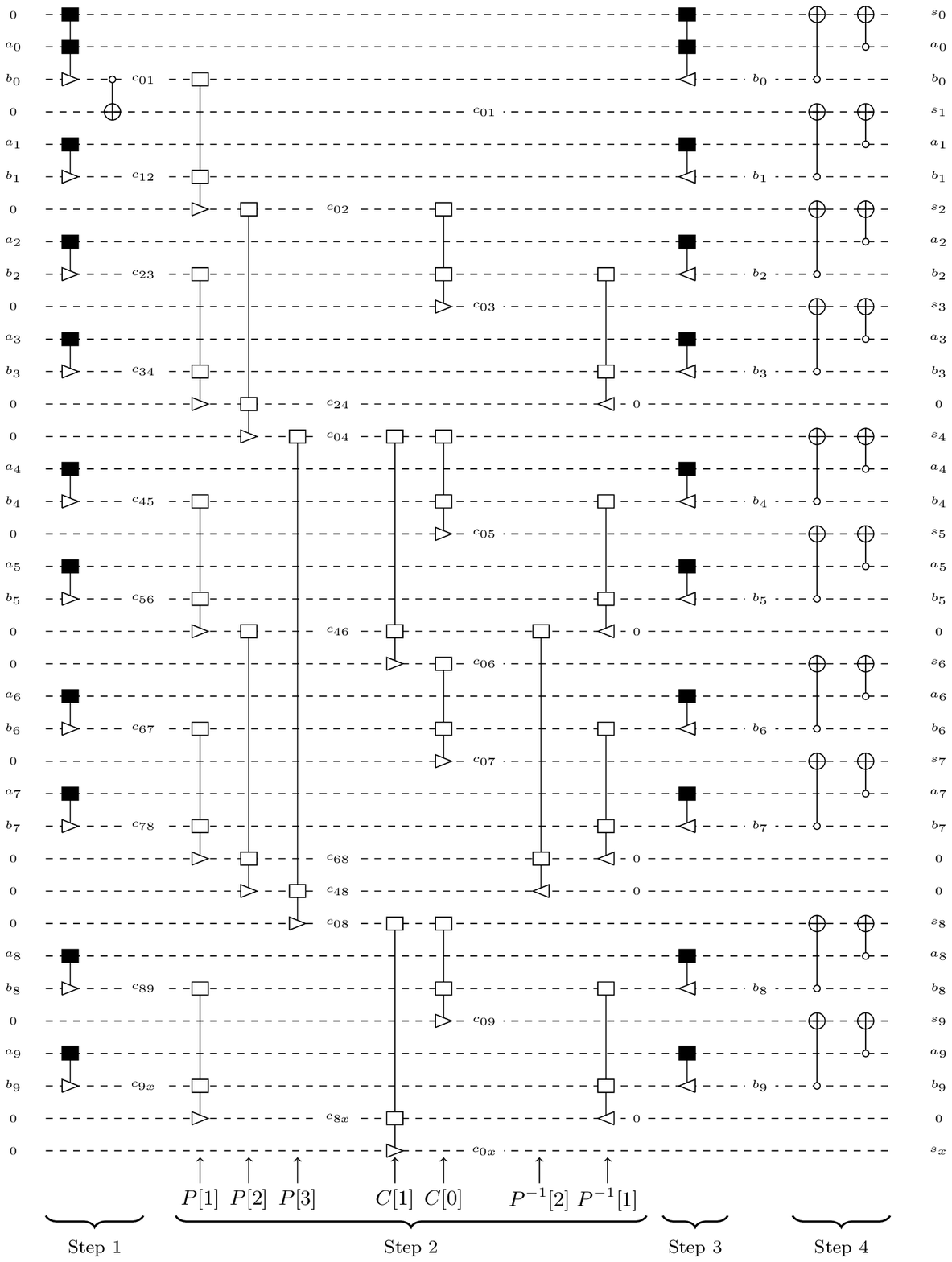}
\fcaption{Out-of-place carry look-ahead adder}\label{fig:out-of-place}
\end{figure}

We represent $\AdjC_0, \; \AdjC$ and $\M$ as shown in Figure \ref{fig:circuitrep}. Their inverses are represented by the same circuit with \begin{tikzpicture}[scale = 0.5]\draw (0,-0.2) -- (0,0.2) -- (0.5,0) -- (0,-0.2);\end{tikzpicture} replaced by \begin{tikzpicture}[scale = 0.5]\draw (0.5,-0.2) -- (0.5,0.2) -- (0,0) -- (0.5,-0.2);\end{tikzpicture}. Also a black rectangle means the content will be changed after the application of the relevant gate, while a blank rectangle means the content remains the same. An an illustration, we give a complete out-of-place circuit for adding two $10$-qutrit numbers in Figure \ref{fig:out-of-place}, where we use $x$ to stand for $10$, and $c_{ij}$ is the carry status indicator $C[i,j]$. From Figure \ref{fig:out-of-place}, it is clear that the $C[0]$-round and $P^{-1}[2]$-round can be parallelized since the gates in these two rounds act on different wires. One can also verify the cost: the number of ancillas is $n-\omega(n) - \fl{\log\,n} = 5$, the depth of the circuit is $\fl{\log\,n} + \fl{\log\,\frac{n}{3}} +6 = 10$, and the complexity is $5n-2\omega(n) - 2\fl{\log\,n}  -1 = 39$.

\begin{figure}
\centering
\begin{tikzpicture}[scale = 0.5]
\DashLine{-0.5}{0}{1.5}
\DashLine{-0.5}{1}{1.5}
\DashLine{-0.5}{2}{1.5}

\DashLine{2.5}{0}{1.5}
\DashLine{2.5}{1}{1.5}

\DashLine{5.5}{0}{1.5}
\DashLine{5.5}{1}{1.5}
\DashLine{5.5}{2}{1.5}

\CarryZero{0}{0}{1}{2}
\Carry{3}{0}{1}
\Merge{6}{0}{1}{2}
\draw (0.25,-1) node{${\sf AdjC}_0$};
\draw (3.25,-1) node{${\sf AdjC}$};
\draw (6.25,-1) node{$\mathcal{M}$};
\end{tikzpicture}
\fcaption{Circuit glyphs for $\AdjC_0, \; \AdjC$ and $\M$. The inverse gates $\AdjC_0^{-1}, \; \AdjC^{-1}$ and $\M^{-1}$ are represented by mirror images of these glyphs. }\label{fig:circuitrep}
\end{figure}

\subsubsection{In-place Adder}
\label{subsubsec:in-place}
The idea of in-place adder is also generalized from that in \cite{draper2006logarithmic}. Let $\bar{2}$ be the $n$-trit number with all $2\,'$s, namely $\bar{2} = 3^n-1$. When no confusion arises, we make no distinction between a number and its trit representation. For two $n$-trit numbers $a,b$, denote by $a \oplus b$ the number obtained by trit-wise summation modulo $3$, and denote by $a'$ the number obtained by replacing every trit $a_i$ by $2-a_i$. Thus, the following equations hold:

\begin{equation*}
a \oplus a' = \bar{2} \text{ and } a + a' = 3^n-1.
\end{equation*}

Let $c= c_0\cdots c_{n-1}$ be the sequence of the $n$ low carry trits for $a$ and $b$, and let $s$ be the $n$ low trits of $a+b$. Then we have
\begin{equation*}
s = a+b \;(\text{mod }3^n) \text{ and } s = a \oplus b \oplus c.
\end{equation*}

Also note that $s'+a = 3^n-1-s+a = 3^n-1-b = b' $ (mod $3^n$).

Let $d = d_0 \cdots d_{n-1}$ be the $n$ low carry trits resulting from adding $s'$ and $a$. Then, $s' \oplus a \oplus d = b'$, and thus we have,

\begin{align*}
\bar{2} \oplus a \oplus b \oplus d  \quad = &\quad  s  \oplus s' \oplus a \oplus b \oplus d \\
                                     = &\quad  s \oplus b' \oplus b \\
                                     = & \quad \bar{2} \oplus a \oplus b \oplus c. \\
\end{align*}

Therefore, $c = d$, i.e., the $n$ low carry trits for $a,b$ are the same as those for $s',a$. We will use this property to implement the in-place adder.

For $0 \leq i \leq n-1$, let $A_i, B_i$ be the working registers initialized with $a_i,b_i$, respectively. We will need $2n-\omega(n) - \fl{\log\,n}$ ancillas, $n$ of which are denoted by $Z_0, Z_1, \cdots, Z_{n-1}$ and the rest are $X_i\,'$s. Let $Z_{n}$ be the working register which will store the high trit of $a+b$. All ancillas start with $0$.

\vspace{0.5cm}
\textbf{In-place Procedure:}
\begin{enumerate}
  \item As described in Out-of-place Procedure Step $1$ through $3$, compute all the carry trits $C[0,j]$ into $Z_{j}, j = 0, \cdots, n$. The ancillas $X_i\,'$s and working registers $A_i, B_i$ are all returned to their initial configuration at the end of the process. This has a circuit depth of $\fl{\log\,n} + \fl{\log\,\frac{n}{3}} +6$, with the complexity of $5n-2\omega(n) - 2\fl{\log\,n}  +1$.
  \item For $0 \leq i \leq n-1$, let $B_i = B_i \oplus A_i \oplus Z_i$, namely, the register $B_i\,'$s will store the $n$ low trits of the sum $a+b$. This can be done by $2n$ $\SUM$ gates.
  \item Now we want to erase the $n$ carry trits $C[0,i] = c_i$, $i=0, \cdots, n-1$. For $0 \leq i \leq n-2$, let $B_i = 2 - B_i$. This can be achieved by $n-1$ $S_{0,2}$ gates.
  \item Apply the inverse of the Out-of-place Procedure Step $1$ through $3$ on the registers $A_i, B_i$ for $0 \leq i \leq n-2$ to erase the carry trits $c_j$ stored in $Z_j, j = 0, \cdots, n-1$.
  \item For $0 \leq i \leq n-2$, let $B_i = 2 - B_i$. Again this can be done by $n-1$ $S_{0,2}$ gates.
\end{enumerate}
\vspace{0.5cm}

Tracing the cost of the circuit above, we see that the in-place adder has a depth of $\fl{\log\,n} + \fl{\log\,\frac{n}{3}} + \fl{\log\,(n-1)} + \fl{\log\,\frac{n-1}{3}} +12$, and its complexity is $10n-2\omega(n) - 2\fl{\log\,n}  -2\omega(n-1) - 2\fl{\log\,(n-1)}  -3$. Moreover, the number of ancillas required is $2n-\omega(n) - \fl{\log\,n}$.

\begin{figure}
\includegraphics[width = 1\textwidth]{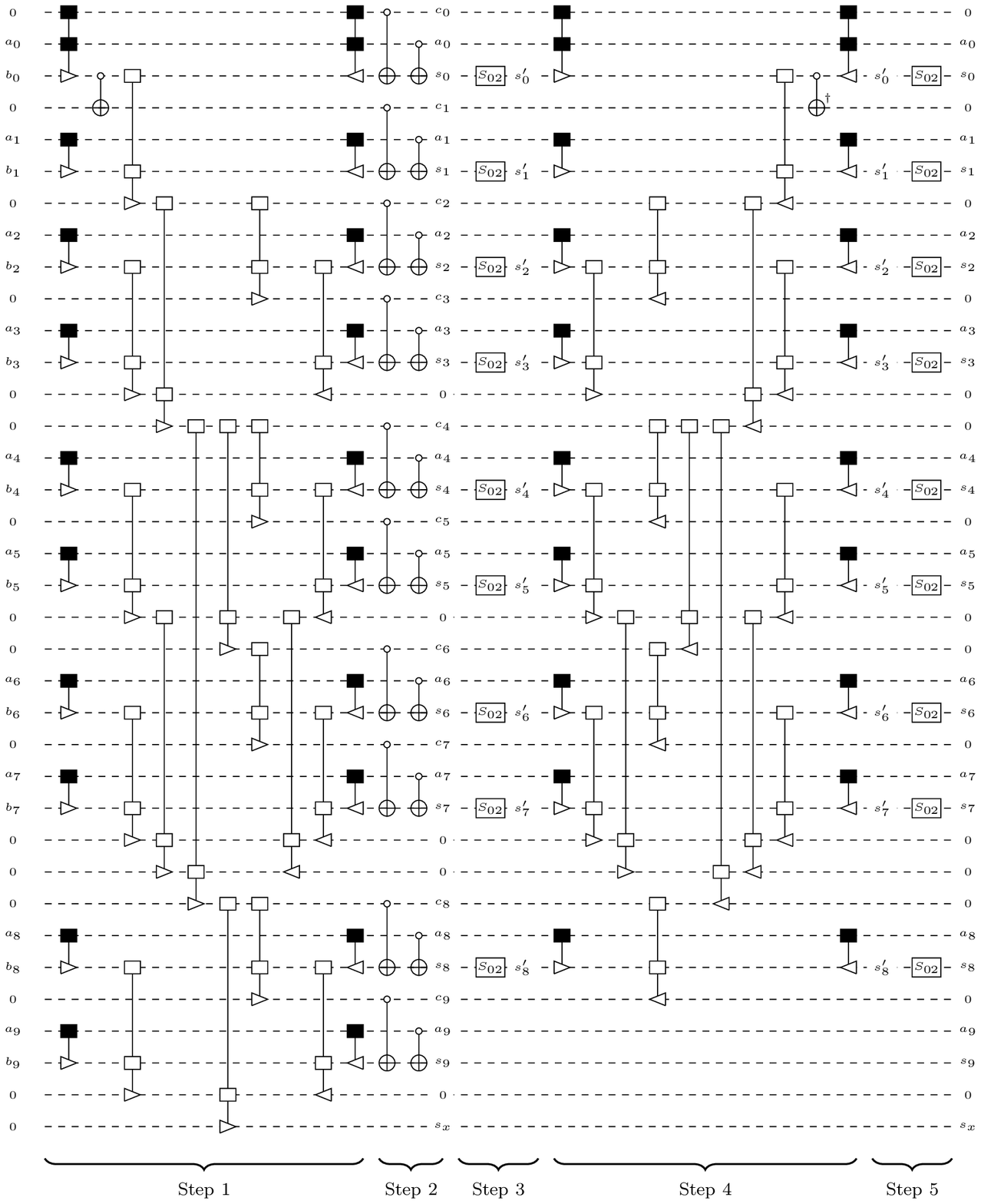}
\fcaption{In-place carry look-ahead adder}\label{fig:in-place}
\end{figure}

Figure \ref{fig:in-place} gives a complete circuit of in-place adder for $n=10$. See Figure \ref{fig:circuitrep} and the last paragraph in Section \ref{subsubsec:out-of-place} for the explanations of notations used in the circuit.

\section{Extensions}
\label{sec:extensions}

In this section, we give various extensions based on the modified ripple-carry adder and the carry look-ahead adder, including addition modulo $3^n$, subtraction, and comparison.

\subsection{Addition Mod $3^n$ }
 \label{subsec:mod3n}
To add two $n$-qutrit numbers modulo $3^n$, we simply do not compute the the high carry trit $c_n$.

In the ripple-carry adder (see Figure \ref{fig:rippleadder}), it suffices to remove the circuit $C$, $\SUM$, $C^{-1}$ in the middle, and the last qutrit on the bottom. Thus in total we need $1$ ancilla, $2(n-1)$ Carry gates, and $2n-1$ $\SUM$ gates, and the depth of the circuit is $4(n-1)$.

In the out-of-place carry look-ahead adder, we run the circuit as described in Out-of-place Procedure in Section \ref{subsubsec:out-of-place}. However, in the first three steps of the procedure, we restrict the inputs to the $n-1$ low trits of $a$ and $b$, namely, $a_0,\cdots,a_{n-2}, b_0,\cdots,b_{n-2}$, since there is no need to compute $c_{n}$. Of course, in the last step we still need to compute the modulo summation $a_i \oplus b_i \oplus c_i$ for all $0 \leq i \leq n-1$. Thus the out-of-place modulo adder uses $n-1-\omega(n-1) - \fl{\log\,(n-1)}$ ancillas, and has a circuit depth of $\fl{\log\,(n-1)} + \fl{\log\,\frac{n-1}{3}} +6$, with  complexity $5(n-1)-2\omega(n-1) - 2\fl{\log\,(n-1)}  +1$.

Similarly, for the in-place carry look-ahead modulo $3^n$ adder, we run exactly the same circuit as the In-place Procedure in Section \ref{subsubsec:in-place}, except in Step $1$ where we again restrict the inputs only to the $n-1$ low trits of $a$ and $b$. It is direct to total the cost of the circuit. It has a depth of $2(\fl{\log(n-1)} + \fl{\log\frac{n-1}{3}}+6)$, with the complexity of $2(5(n-1)-2\omega(n-1) - 2\fl{\log\,(n-1)} +1)$. The number of ancillas required is $2(n-1)-\omega(n-1) - \fl{\log\,(n-1)}$.


 \subsection{Subtraction}
 \label{subsec:subtraction}
 To compute $a-b$ for two $n$-trit numbers $a,b$, first convert $a$ to $a'$, then compute $a'+b$, and eventually convert $a'+b$ to $(a'+b)'$. Note that $a'$ is the $n$-trit number obtained by replacing each $a_i$ by $2-a_i$, namely, $a' = 3^n-1-a$. Thus we have,
 \begin{equation*}
 (a'+b)' = (3^n-1-a+b)'= 3^n-1- (3^n-1-a+b) = a-b.
 \end{equation*}

Changing $a$ to $a'$ costs $n$ Clifford gate $S_{0,2}$. Therefore, the circuit for subtraction has the same depth and complexity as the regular the adder.

 \subsection{Comparison}
 \label{subsec:comparison}

Given the circuit for subtraction, it is straightforward to compare two numbers $a$ and $b$. Actually, there is a circuit for the comparison of $a,b$ with  smaller complexity than that of subtraction since we only need to know the high trit of $a-b$. Let $a' = 3^n-1-a$, then $a - b \geq 0$ if and only if the high trit of $a' + b$ is $0$.

In the ripple-carry adder, we convert $a$ to $a'$ and use the Carry gate $C$ to compute all the carry trits $c_1, \cdots, c_{n}$ for $a' + b$. After copying $c_n$ to the register storing the result of the comparison, we undo all the $C\,'$s and convert $a'$ back to $a$. The circuit thus requires $1$ ancilla, $2n$ Carry gate $C$, $1$ $\SUM$ gate, $2n$ $S_{0,2}$, and has a depth of $4n$.


In the carry look-ahead adder, again we first convert $a$ to $a'$. To compute $a'+b$, the circuit sequentially generates all the carry status indicators $C[i,j]\,'$s. However, since we only care about the high trit $c_n = C[0,n]$, we can design a more efficient circuit to implement the comparison.

Recall from Section \ref{subsec:lookahead} that in $P$ process we have obtained all the carry status indicators of the form $C[2^t m,2^t(m+1)]$, and in particular, any $C[0, 2^k]$ is of this form. Therefore, if $n=2^k$ for some $k$, then $c_n$ is obtained at the end of $P$ process. At this moment, there is no need to go through the $C$ process. Instead, we copy $c_n$ into the register storing the result, and undo the $P$ process. In general, let $k = \ceil{\log\,n}$, then we can just pad $a$ and $b$ by adding zeros in the front to make them $2^k$-trit numbers, and use the circuit described above to compare $a$ and $b$. We still call the $2^k$-trit numbers $a$ and $b$. For $0 \leq i \leq n-1$, let $A_i = a_i, B_i = b_i$ be the working registers, and let $R$ the register which will store the result of the comparison. We also need $2^k + 2(2^k-n)$ ancillas, among which $2(2^k-n)$ are used to hold the extra zeros in from of $a$ and $b$, one is denoted by $Z_0$ as the ancilla to the $\AdjC_0$ circuit, and the rest are denoted by $X_i\,'$s.

Note that after padding $a$ and $b$ with zeros, the carry status indicators $C[i,j]\,'$s, $n \leq i < j\leq 2^k$, are known before the compilation, thus we can store their values in the registers and there is no need to recompute them later.

\vspace{0.5cm}
\textbf{Carry Look-ahead Comparison:}
\begin{enumerate}
  \item Convert $a$ to $a'$. This requires $2^k$ $S_{0,2}$ gates.
  \item For $ 0 < i \leq n-1$, run the circuit $\AdjC$ on $A_i, B_i$, which outputs $C[i,i+1]$ to $B_i$. Run $\AdjC_0$ on $A_0,B_0$, and $Z_0$ with $Z_0$ as the ancilla, which outputs $C[0,1]$ to $B_0$. The circuit has a depth of $2$, and it consist of $n-1$ $\AdjC$ and $1$ $\AdjC_0$.
  \item Perform the $P$ process in Section \ref{subsec:lookahead} to compute all the $C[2^t m,2^t(m+1)]$ that are not known before compilation into the ancillary registers $X_i$. Note that here since we don't have the $Z_i$ registers, all the $C[0,2^m]\,'$s are also written to the $X_i$ registers. The depth of the circuit is $k$, and the complexity is $2^k-\omega(2^k) - (2^k-n-\omega(2^k-n)) = n + \omega(2^k-n) -1$.
  \item Copy $c_{2^k}$ to the result register $R$.
  \item Undo Step $3$.
  \item Undo Step $2$.
  \item Undo Step $1$.
\end{enumerate}

Therefore, the total depth of the circuit above is $2k+4 = 2\ceil{\log\,n} + 4$, and it has the complexity of $4n + 2\omega(2^k-n) = 4n + 2\omega(2^{\ceil{\log\,n}}-n)$. The number of ancillas used is $3 \cdot 2^{\ceil{\log\,n}} -2n$.


\section{Techniques for Constructing Quantum Gate Decompositions}
\label{sec:techniques}
In previous sections, we developed a system of ternary arithmetic with the focus on two types of quantum ternary adders. The building blocks of these circuits include the Carry circuit $C$, the circuits $\AdjC, \AdjC_0$ computing carry status indicators, and the {\it merging} formula $\M$. Moreover, the non-Clifford gates used in these four circuits are $\TwoSwap, \CU{S_{0,1}}, \CX,$ and $\CSUM$.

In this section, we show that it suffices to have $\CX$ along with Clifford gates to produce the other three non-Clifford gates exactly. The key technique involved is to analyze the algebraic expressions of these gates. In Section \ref{subsec:construction2}, it is proven that $\CX$ and $\Horner$ are equivalent up to Clifford gates, and that all other non-Clifford gates can be obtained from $\CX$. In Section \ref{subsec:super}, we introduce a universal gate set called \super, which is a qutrit analog of the qubit Clifford $+$ $\frac{\pi}{8}$-gate. We then illustrate in Section \ref{subsec:construction1} that $\CX$ and $\Horner$ can both be implemented exactly over \super. Therefore, with the \super, the ternary circuits for arithmetic can be realized exactly.

\subsection{Construction of Reversible Gates from Polynomial Expressions}
 \label{subsec:construction2}

 Let $\F_3$ be the field with three elements $\{0,1,2\}$. Then any $n$-qutrit reversible gate can be represented as a map $\F_3^n \mapsto \F_3^n$, or a sequence of $n$ functions $\F_3^n \mapsto \F_3$, if one identifies each $\ket{i}$ with $i$, $i=0,1,2$. We will see  that reversible gates have polynomial representations and these polynomial representations provide hints to construct one reversible gate from another.

Note that $0^2 = 0, 1^2 = 2^2 = 1 \, ({\rm mod} \; 3)$, and thus $\delta_{i,0} = 1-i^2 \, ({\rm mod} \; 3)$. By default, arithmetic within a ket is taken modulo $3$. The following is a list of polynomial expressions of some non-Clifford gates.

\begin{itemize}
\item $\SUM = \bigwedge(X): \ket{i,j} \mapsto \ket{i,i+j}$;

\item $\HardU{0}{X}: \ket{i,j} \mapsto \ket{i,j + \delta_{i,0}} = \ket{i, j - i^2 + 1}$; 
\item $\Horner$$ := \SoftU{\SoftU{X}}:\ket{i,j,k} \mapsto \ket{i,j,ij+k}$;
\item $\HardU{0}{\SUM}:\ket{i,j,k} \mapsto \ket{i,j,k+(1-i^2)j}$.
\end{itemize}

 The above list shows that if a qutrit works as a soft control, then it contributes a linear factor in the expression of the target qutrit, while a hard control qutrit contributes a quadratic factor.

Define $C'(X): \ket{i,j} \mapsto \ket{i,j+i^2}$. Thus, $C'(X) = (I \otimes X){\HardU{0}{X}}^{-1}$ is equivalent to $\CX$. We will use $C'(X)$ below for the construction of other gates.

The relation between the expressions of $\Horner$ and $\C'(X)$ resembles that of a bilinear form and a quadratic form, which are equivalent. This suggests that $\Horner$ and $C'(X)$ are also equivalent. Indeed, the following diagrams give a construction of one from another.

\begin{itemize}
  \item {\small{implementation of $\Horner$ gate in terms of $C'(X):\,$ $\ket{i,j,k}$ $\overset{\SUM_{1,2}}{\longrightarrow}$ $\ket{i,i+j,k}$ $\overset{C'(X)_{2,3}^{-1}}{\longrightarrow}$ $\ket{i,i+j,k-(i+j)^2}$ $\overset{\SUM_{1,2}^{-1}}{\longrightarrow}$ $\ket{i,j,k-i^2-j^2+ij}$ $\overset{C'(X)_{1,3}}{\longrightarrow}$ $\ket{i,j,k-j^2+ij}$ $\overset{C'(X)_{2,3}}{\longrightarrow}$ $\ket{i,j,k+ij}$.}}
  \item  {\small{implementation of $C'(X)_{1,2}$ gate in terms of $\Horner:\, $ $\ket{i,j,k}$ $\overset{\SUM_{1,3}}{\longrightarrow}$ $\ket{i,j,i+k}$ $\overset{\Horner_{1,3,2}}{\longrightarrow}$ $\ket{i,j+i^2+ik, i+k}$ $\overset{\SUM_{1,3}^{-1}}{\longrightarrow}$ $\ket{i,j+i^2+ik,k}$ $\overset{\Horner_{1,3,2}^{-1}}{\longrightarrow}$ $\ket{i,j+i^2,k}$.}}
\end{itemize}

Note that in the construction of $2$-qutrit $C'(X)$, we made use of a third qutrit, but that qutrit does not have to be clean, namely it could have arbitrary state.

Similarly, $C'(X)$ is enough to construct $\CSUM$:

$\HardU{0}{\SUM}$: $\ket{i,j,k}$ $\overset{C'(X)_{1,2}}{\longrightarrow}$ $\ket{i,i^2+j,k}$ $\overset{C'(X)_{2,3}}{\longrightarrow}$ $\ket{i,i^2+j,k+(i^2+j)^2}$ $\overset{C'(X)_{1,2}^{-1}}{\longrightarrow}$ $\ket{i,j,k+i^2+j^2-i^2j}$ $\overset{C'(X)_{1,3}^{-1}}{\longrightarrow}$ $\ket{i,j,k+j^2-i^2j}$ $\overset{C'(X)_{2,3}^{-1}}{\longrightarrow}$ $\ket{i,j,k-i^2j}$ $\overset{\SUM_{2,3}}{\longrightarrow}$ $\ket{i,j,k+(1-i^2)j}.$

To implement $\CU{S_{0,1}}$ and $S_{00,22}$, notice that the circuit in Figure \ref{fig:S0110} realizes $S_{01,10}$, and moreover we have:

\begin{itemize}
 \item $S_{00,22} = \SUM^{-1}(X^{-1} \otimes I)S_{01,10}(X \otimes I)\SUM$.
 \item $\HardU{0}{S_{0,1}}=$ $ \SUM_{2,1}^{-1}(X^{-1} \otimes X^{-1}) S_{00,22} (X \otimes X)\SUM_{2,1}$.
\end{itemize}

 \begin{figure}
 \centering
\begin{tikzpicture}[scale = 0.5]
\begin{scope}
  \drawCX{0}{0}{1};
  \draw (0.8,0.8) node{\tiny{$2$}};
  \drawCXPrime{1.5}{0}{1};
  \draw (2.3,-0.2) node{\tiny{$2$}};
  \drawCX{3}{0}{1};
  \draw (3.8,0.8) node{\tiny{$2$}};
  \drawCXPrime{4.5}{0}{1};
  \draw (5.3,-0.2) node{\tiny{$2$}};
  \drawCX{6}{0}{1};
  \draw (6.8,0.8) node{\tiny{$2$}};
  \drawTwoU{8}{0}{1}{\tiny{SWAP}};
\end{scope}
\end{tikzpicture}
\fcaption{A circuit for $S_{01,10}$}\label{fig:S0110}
\end{figure}

 \subsection{Supermetaplectic Basis}
 \label{subsec:super}
Recall from Section \ref{sec:background} that $\C$ is the qutrit Clifford group generated by $H, Q, X$, and $\SUM$. Some other gates in $\C$ are $Z$ and $\bigwedge(Z)$, where $Z = \diag(1,\zeta_3,\zeta_3^2)$, and $\bigwedge(Z) = (I \otimes H)\SUM(I \otimes H^{-1})$. It can be directly verified that $\bigwedge(Z)$ has the following expression:
$$\bigwedge(Z): \ket{i,j} \mapsto \zeta_3^{ij}\ket{i,j}.$$
In \cite{cui2015universal}, it has been established that the multi-qutrit \emph{metaplectic} gate set $\C$ $+$ $\diag(1,1,-1)$ or equivalently $\C$ $+$ $\diag(1,\zeta_6, \zeta_6^2)$ was universal for quantum computation  in the sense that any multi-qutrit unitary operator can be approximated to any given precision by a circuit over that gate set.
We conjecture that the metaplectic gate set is not universal for \emph{exact} reversible computation, i.e. it seems that the subgroup of reversible classical gates that can be represented exactly by metaplectic circuits is rather thin. 
In order to ensure exact representation of the reversible gates over a relatively simple multi-qutrit basis, we expand the basis by adding essentially the ``cubic root'' of the $Z$ gate to it. To
this end we increase the order of the root of unity used in defining the non-Clifford diagonal gate, and define $P_9$ as the $1$-qutrit diagonal gate $\diag(\zeta_9^{-1},1,\zeta_9)$.\footnote{This is the the distillable gate denoted $M_3^{\dagger}$ in \cite{campbell2012magic}. }

 \begin{defin}
 The gate set $\C$ $+$ $P_9$ is called $\super$.
 \end{defin}
%
%


Since the $P_9$ gate is non-Clifford, this basis is universal for quantum computation.
The supermeta\-plectic basis resembles the qubit Clifford $+$ $T$ basis in several aspects. Firstly, we show in Section \ref{subsec:construction1} that all the reversible gates can be constructed exactly over the $\super$. Secondly, the $P_9$ gate is a fundamental diagonal gate in the third level of the Clifford hierarchy \cite{howard2012qudit}. Lastly, it was shown in \cite{campbell2012magic} that $P_9$ can be obtained by magic state distillation.
 %

 \subsection{Construction of Diagonal Gates from Polynomial Expressions}
 \label{subsec:construction1}

 We continue exploring the use of polynomial expressions in constructing new quantum gates.

The group of reversible gates in $\C$ is generated by $\SUM, X, S_{1,2}$. More precisely, it is described by the following proposition.

  \begin{prop}\label{prop:reversible}
    $\{S_{12},  X,\SUM \}$ generate a maximal subgroup, which is isomorphic to $\simeq \GL(n,\mathbb{F}_3) \rtimes \mathbb{F}_3^n$, of the group of reversible gates for any number $n$ of qutrits.
  \end{prop}
   \vspace*{12pt}
\noindent
{\bf Proof:} See Appendix \ref{app:reversible}. \square

The statement in Proposition \ref{prop:reversible} for the case $n=2$ was also proved in \cite{bocharov2015efficient}.

By the proof of Proposition \ref{prop:reversible}, the correspondence between $\GL(n,\mathbb{F}_3) \rtimes \mathbb{F}_3^n$ and the group generated by $\{S_{12},  X,\SUM \}$ is as follows:

Given a pair $(A,v) \in \GL(n,\mathbb{F}_3) \rtimes \mathbb{F}_3^n$, where $A = (a_{ij})_{1 \leq i,j \leq n}, v = (v_i)_{1 \leq i \leq n}$, then the reversible $n$-qutrit gate corresponding to it maps $\ket{x}$, for any computational basis element $\ket{x} = \ket{x_1, \cdots, x_n}$, to $\ket{A.x+v}$. Moreover, any reversible gate of this form is generated by $\{S_{12},  X,\SUM \}$.

 A function $f: \F_3^n \mapsto \F_3$ is called affine linear if $f(x_1,\cdots,x_n) = a_1 x_1 + \cdots + a_n x_n + b$, where $a_1, \cdots, a_n, b \in \F_3$. A reversible $n$-qutrit gate can be viewed as an $n$-tuple of functions: $\ket{x} \mapsto \ket{f_1(x), \cdots, f_n(x)}$, where we call $f_i$ the coordinates of the gate. Then the above argument shows that a reversible $n$-qutrit gate is generated by $\{S_{12},  X,\SUM \}$ if and only if all of its coordinates are affine linear functions. Let $\mathcal{F}_n$ be the set of all affine linear functions from $\F_3^n$ to $\F_3$.

 Let $\D$ be the group generated by the reversible gates in $\C$, together with the diagonal gates $\bigwedge(Z)$ and $P_9$. We give a technique to characterize all the diagonal gates in $\D$.

 By Proposition \ref{prop:reversible} and the argument above, the reversible gates in $\D$ can change the basis element $\ket{x}$ to any element of the form $\ket{f_1(x), \cdots, f_n(x)}$, where $f_i$ is an affine linear function $\F_3^n$ to $\F_3$. The action of $\bigwedge(Z)$ and $P_9$ will contribute a scalar to the basis element. Thus the most general $n$-qutrit diagonal gate in $\D$ has the form:
%
%

\begin{equation}\label{equ:form}
\ket{i_1,i_2,\cdots, i_n} \mapsto \zeta_9^{\sum\limits_{f \in \mathcal{F}_n}A_f f(i_1,\cdots,i_n)}\zeta_3^{\sum\limits_{f,g \in \mathcal{F}_n} B_{f,g} f(i_1,\cdots, i_n)g(i_1,\cdots,i_n)}\ket{i_1,i_2,\cdots, i_n},
\end{equation}
where $A_f, B_{f,g}$ are integer parameters. Notice that the affine linear functions $f$ and $g$ take values in $\F_3$, while $A_f, B_{f,g}$ take values in $\Z$. We have to evaluate $f,g$ first in $\{0,1,2\}$, then multiply by $A_{f}, B_{f,g}$ inside $\Z$. This is critical for the term $\zeta_9$.

%
%
%
%
%
%

As an application, we show that $\bigwedge(\bigwedge(Z))$ and $\HardU{2}{Z}$ are both contained in $\D$. The expressions of relevant gates are given below.

\begin{itemize}
 \item $ \bigwedge(Z)|i,j\rangle = \zeta_3^{ij}|i,j\rangle, P_9|i\rangle = \zeta_9^{i}|i\rangle,$
 \item $X |i\rangle = |i+1\rangle, S_{1,2}|i\rangle = |2i\rangle, \SUM|i,j\rangle =|i,i+j\rangle$.
\item $\SoftU{\SoftU{Z}}: \ket{i,j,k} \mapsto \zeta_3^{ijk}\ket{i,j,k}$.
 \item $\HardU{2}{Z}: \ket{i,j} \mapsto \zeta_3^{j\delta_{i,2}}\ket{i,j}$.
\end{itemize}

For $n=3$, the coefficient in Formula \ref{equ:form} can be written as:

  \begin{equation}
  L(i,j,k) = \zeta_9^{\sum\limits_{a,b,c,d=0}^{2} A_{a,b,c,d} (ai+bj+ck+d)} \zeta_3^{Bij+C jk+Dik}, \quad i,j,k \in \F_3,
  \end{equation}
   where $A_{a,b,c,d}, B, C, D$ are integer parameters \footnote{Actually there are also terms $i^2, j^2, k^2$ on the exponent of $\zeta_3$, but it is direct to see that $\zeta_3^{i^2} = \zeta_9^{(2i \, {\rm mod} \; 3) - ((2-i) \, {\rm mod} \; 3)}$ up to a global phase, so the square terms can be absorbed into the $\zeta_9$ terms.}. Again $ai+bj+ck+d$ is assumed to be taken modulo $3$.

  To construct $\bigwedge(\bigwedge(Z))$, set $L(i,j,k) = \zeta_3^{ijk}$. Since $\zeta_9 = \zeta_3^3$, we get the equation:

  \begin{equation}
  \Equ(i,j,k): \sum\limits_{a,b,c,d} A_{a,b,c,d} (ai+bj+ck+d) + 3(Bij+Cjk+Dik) = 3ijk \,(\text{ mod }9), \quad i,j,k \in \F_3.
  \end{equation}

  The set $\{\Equ(i,j,k): i,j,k \in \F_3\}$ is a system of $27$ linear equations in the variables $A_{a,b,c,d}, B, C,$ and $D$. Thus there is an efficient way to find the solutions, if any.

  By direct calculations, one solution to the above system of equations is:

\begin{equation}\label{equ:solution}
\zeta_3^{ijk} = \zeta_9^{(1 + 2 i + j + k)+2(1 + 2 i + j + 2 k)+6(2 + 2 i + j + 2 k)+2(1 + 2 i + 2 j + k)+6(2 + 2 i + 2 j + k)+4(1 + 2 i + 2 j + 2 k)+6(2 + 2 i + 2 j + 2 k)},
\end{equation}
where the terms on the exponent within each parenthesis is taken modulo $3$.

 In light of the solution in Equation \ref{equ:solution}, it is not hard to create a circuit realizing $\SoftU{\SoftU{Z}}$. Explicitly, this is given in Figure \ref{fig:wedge2Z}.

\begin{figure}
\centering
  \setlength{\unitlength}{0.025in}
  \begin{picture}(170,35)(0,-3)
    \put(0,0){\line(1,0){8}}
    \put(0,10){\line(1,0){10}}
    \put(0,20){\line(1,0){10}}

    \put(8,-3){\framebox(8,6){{\tiny{$S_{1,2}$}}}}
    \put(16,0){\line(1,0){4}}
    \put(10,10){\line(1,0){10}}
    \put(10,20){\line(1,0){10}}

    \put(20,-3){\framebox(6,6){{\tiny{$X$}}}}
    \put(26,0){\line(1,0){4}}
    \put(20,10){\line(1,0){10}}
    \put(20,20){\line(1,0){10}}

    \put(30,0){\line(1,0){10}}
    \put(30,10){\line(1,0){10}}
    \put(30,20){\line(1,0){10}}
    \put(33,0){\circle{6}}
    \put(33,10){\line(0,-1){13}}
    \put(33,10){\color{white}\circle*{2}}
    \put(33,10){\circle{2}}

    \put(40,0){\line(1,0){10}}
    \put(40,10){\line(1,0){10}}
    \put(40,20){\line(1,0){10}}
    \put(43,0){\circle{6}}
    \put(43,20){\line(0,-1){23}}
    \put(43,20){\color{white}\circle*{2}}
    \put(43,20){\circle{2}}

    \put(50,0){\line(1,0){10}}
    \put(50,10){\line(1,0){10}}
    \put(50,20){\line(1,0){10}}
    \put(53,0){\line(0,1){23}}
    \put(53,20){\circle{6}}
    \put(53,0){\color{white}\circle*{2}}
    \put(53,0){\circle{2}}

    \put(60,0){\line(1,0){10}}
    \put(60,10){\line(1,0){10}}
    \put(60,20){\line(1,0){10}}
    \put(63,0){\line(0,1){13}}
    \put(63,10){\circle{6}}
    \put(63,0){\color{white}\circle*{2}}
    \put(63,0){\circle{2}}

    \put(70,-3){\framebox(6,6){{\tiny{$P_9$}}}}
    \put(76,0){\line(1,0){4}}
    \put(70,7){\framebox(6,6){{\tiny{$P_9^{\dag}$}}}}
    \put(76,10){\line(1,0){4}}
    \put(70,17){\framebox(6,6){{\tiny{$P_9^{\dag}$}}}}
    \put(76,20){\line(1,0){4}}


    \put(80,0){\line(1,0){10}}
    \put(80,10){\line(1,0){10}}
    \put(80,20){\line(1,0){10}}
    \put(83,10){\circle{6}}
    \put(86,12){{\tiny{$\dag$}}}
    \put(83,0){\line(0,1){13}}
    \put(83,0){\color{white}\circle*{2}}
    \put(83,0){\circle{2}}

    \put(90,0){\line(1,0){10}}
    \put(90,10){\line(1,0){10}}
    \put(90,20){\line(1,0){10}}
    \put(93,20){\circle{6}}
    \put(93,10){\line(0,1){13}}
    \put(93,10){\color{white}\circle*{2}}
    \put(93,10){\circle{2}}

    \put(100,17){\framebox(6,6){{\tiny{$P_9$}}}}
    \put(106,20){\line(1,0){4}}
    \put(100,0){\line(1,0){10}}
    \put(100,10){\line(1,0){10}}

    \put(110,17){\framebox(6,6){{\tiny{$X$}}}}
    \put(116,20){\line(1,0){3}}
    \put(110,0){\line(1,0){10}}
    \put(110,10){\line(1,0){10}}

    \put(119,17){\framebox(8,6){{\tiny{$S_{1,2}$}}}}
    \put(127,20){\line(1,0){3}}
    \put(120,0){\line(1,0){10}}
    \put(120,10){\line(1,0){10}}

    \put(130,0){\line(1,0){10}}
    \put(130,10){\line(1,0){10}}
    \put(130,20){\line(1,0){10}}
    \put(133,0){\circle{6}}
    \put(136,2){{\tiny{$\dag$}}}
    \put(133,20){\line(0,-1){23}}
    \put(133,20){\circle{2}}
    \put(133,20){\color{white}\circle*{2}}

    \put(140,0){\line(1,0){10}}
    \put(140,10){\line(1,0){10}}
    \put(140,20){\line(1,0){10}}
    \put(143,20){\circle{6}}
    \put(146,22){{\tiny{$\dag$}}}
    \put(143,10){\line(0,1){13}}
    \put(143,10){\color{white}\circle*{2}}
    \put(143,10){\circle{2}}

    \put(150,0){\line(1,0){10}}
    \put(150,10){\line(1,0){10}}
    \put(150,20){\line(1,0){10}}
    \put(153,20){\circle{6}}
    \put(156,22){{\tiny{$\dag$}}}
    \put(153,0){\line(0,1){23}}
    \put(153,0){\circle{2}}
    \put(153,0){\color{white}\circle*{2}}

    \put(160,17){\framebox(6,6){{\tiny{$X^{\dag}$}}}}
    \put(166,20){\line(1,0){4}}
    \put(160,0){\line(1,0){10}}
    \put(160,10){\line(1,0){10}}
  \end{picture}
  \fcaption{A circuit for $\bigwedge(\bigwedge(Z))$} \label{fig:wedge2Z}
\end{figure}

Similarly, with the same method, we construct a circuit for $\HardU{2}{Z}$. See Figure \ref{fig:C2Z}.

\begin{figure}
\centering
\begin{tikzpicture}[scale = 0.5]
\begin{scope}

  \drawOneU{3}{0}{\tiny{$P_9$}};
  \drawOneU{1.5}{2.5}{\tiny{$Q$}};
  \drawOneU{3}{2.5}{\tiny{$S_{0,2}$}};

  \drawSum{4.5}{0}{2.5};

  \drawOneU{6}{0}{\tiny{$P_9$}};
  \draw (6,2.5) -- (7.5,2.5);

  \drawSum{7.5}{0}{2.5};

  \drawOneU{9}{0}{\tiny{$P_9$}};
  \draw (9,2.5) -- (11,2.5);

  \drawSum{10.5}{0}{2.5};
  \drawOneU{12}{2.5}{\tiny{$S_{0,2}$}};


\end{scope}
\end{tikzpicture}
\fcaption{A circuit for $\HardU{2}{Z}$}\label{fig:C2Z}
\end{figure}


Note that $\SoftU{\SoftU{Z}}, \HardU{2}{Z}$ are related with $\Horner, \HardU{2}{X}$, respectively, by the Clifford gate $H$, namely, we have,
\begin{itemize}
    \item $(I \otimes H) \HardU{2}{X} (I \otimes H^{\dag}) = \HardU{2}{Z} $
    \item $(I \otimes I \otimes H) \Horner (I \otimes I \otimes H^{\dag}) = \SoftU{\SoftU{Z}}$.
  \end{itemize}

Therefore, both $\Horner$ and $\HardU{2}{X}$ can be implemented exactly over \super.

\begin{remark}
\begin{enumerate}
  \item The papers \cite{amy2013meet,AMM:2014} developed a similar framework for the binary case.
  \item If one uses the similar  technique for the qubit Clifford $+$ $T$ gates, namely replacing $(\zeta_9,\zeta_3)$ with $(\zeta_8, -1)$, one obtains a circuit for the Toffoli gate with $T$-depth $3$, which is optimal in the ancilla free scenario.
\end{enumerate}
\end{remark}

\section{Conclusion}
We developed improved ternary circuits for reversible ternary adders of two types: the modified ripple-carry and the carry look-ahead adder. We have also derived solutions for a modulo $3^n$ adder, subtraction and comparison in ternary encoding.
We have offered two levels of abstraction for describing the corresponding ternary circuits: one in terms of reversible reflections of certain types and one in a more uniform language that allows only one non-Clifford gate: either the $C(X): \ket{i,j}\mapsto \ket{i, j + \delta_{i,2} \; {\rm mod} \; 3}$ or the  $P_9=\mbox{diag}(e^{-2 \pi \, i/9}, 1,e^{2 \pi \, i/9})$ gate.

Future circuit synthesis work should entail the design of fully modular adders, circuits for singly- and doubly-controlled adders, as well as optimized circuits for singly- and doubly-controlled additive shifts that would be essential parts of Shor's integer factorization algorithm.

An important theoretical direction of future work would be establishing lower complexity bound for the arithmetic circuits and evaluating the efficiency of designs presented here versus these bounds.

\section{Acknowledgment}
Most of the work in the present paper was done during Summer $2015$ when the second author was interning with Microsoft QuArC Group.

\appendix

\section{Reversible gates generated by $\{S_{12},  X,\SUM \}$ } \label{app:reversible}
 \begin{prop}
    $\{S_{12},  X,\SUM \}$ generate a maximal subgroup, which is isomorphic to $\simeq \GL(n,\mathbb{F}_3) \rtimes \mathbb{F}_3^n$, of the group of reversible gates $($the permutation group$)$ for any number $n$ of qutrits.
  \end{prop}

 \vspace*{12pt}
\noindent
{\bf Proof:} Let $\F_3^n$ be the $n$-dimensional vector space over the finite field $\F_3$. Then there is a one-to-one correspondence between the elements of $\F_3^n$ and the computational basis of the $n$-qutrit space $(\CC^3)^{\otimes n}$. That is, any element $(x_1, \cdots, x_n) \in \F_3^n$ corresponds to the basis element $\ket{x_1, \cdots, x_n}$. Thus any automorphism on $\F_3^n$ induces a permutation on the $n$-qutrit basis, which is a reversible $n$-qutrit gate.

    Let $G = \GL(n,\mathbb{F}_3) \rtimes \mathbb{F}_3^n$, the semidirect product of $\GL(n,\mathbb{F}_3)$ and $\mathbb{F}_3^n$, and let $S_{3^n}$ be the symmetric group on $3^n$ elements, or equivalently the group of reversible gates on $n$ qutrits. We first prove the group generated by $\{S_{12},  X,\SUM \}$ is isomorphic to $G$. As a corollary of applying the O'Nan-Scott Theorem to the classification of maximal subgroups of the symmetric group \cite{scott} \cite{liebeck}, it follows that $G$ is a maximal subgroup of $S_{3^n}$.

 The group $G$ is the affine linear group of degree $n$ over $\F_3$, namely, it consists of all the pairs $(A,v)$, where $A$ is an $n \times n$ invertible group with entries in $\F_3$, and $v$ is a vector in $\F_3^n$. The group $G$ acts on $\F_3^n$ as follows:
$$(A,v).x = A.x + v, \quad (A,v) \in G, x \in \F_3^n$$

Therefore, we get a map $\varphi: G \longrightarrow U(3^n)$, such that $\varphi(A,v)\ket{x} = \ket{Ax+v}$, where $\ket{x}$ is any computational basis vector. This map $\varphi$ is apparently a group homomorphism and injective.

For $1 \leq i \neq j \leq n$, define $A_{ij}, M_i \in \GL(n,\mathbb{F}_3), v_i \in \mathbb{F}_3^n$ as follows.

$A_{ij} = I_n + E_{ji} =
\begin{pmatrix}
1  &        &  & &  & &  \\
   & \ddots &  & &  & &  \\
   &        &1 & &  & &  \\
   &        &  & \ddots & & & \\
   &        &1 &        &1& &  \\
   &        &  &        & & \ddots & \\
   &        &  &        & &        & 1  \\
\end{pmatrix},
\quad
M_i = I_n + E_{ii} = \diag(1, \cdots, 1,2,1,\cdots,1),
\quad
v_i = (0,\cdots,0,1,0,\cdots,0).
$

It is straightforward to check that $\varphi(A_{ij},0) = \SUM_{ij}, \; \varphi(M_i,0) = (S_{1,2})_{i}, \; \varphi(0, v_i) = X_i$, where the subscript of the gate on the right hand side of each expression denotes the qutrits it acts on. For instance, $X_i$ is the $X$ gate acting on the $i$-th qutrit. Therefore, the group generated by $\SUM, X, S_{1,2}$ is isomorphic to the group generated by $A_{ij}, M_i, v_i$, for $1 \leq i \neq j \leq n$.

Clearly all the $v_i\,'$s generate $\F_n^3$ as an additive group. We next prove that $A_{ij}, M_i$ generate the group $\GL(n,\mathbb{F}_3)$.

Let $B_{ij} = M_iA_{ij}A_{ji}^{-1}A_{ij} = I_n - E_{ii} - E_{jj} + E_{ij}+ E_{ji}$, thus $B_{ij}$ swaps the two basis elements $e_i$ and $e_j$. Now given any matrix $A \in \GL(n,\mathbb{F}_3)$, multiplying $A$ on the left by $A_{ij}, B_{ij}$, and $M_i$ constitutes the three types of row operations on $A$, and since $A$ is invertible, it can always be reduced to the identity matrix by row operations. This proves that any matrix in $\GL(n,\mathbb{F}_3)$ can be written as a product of $A_{ij}, B_{ij},$ and $M_i$. Therefore, $\GL(n,\mathbb{F}_3)$ is generated by $A_{ij}, M_i$, and hence $G$ is generated by $A_{ij}, M_i,$ and $v_i$.

Combining the above argument, we showed that the group generated by $\SUM, S_{12}, X$ is isomorphic to $G = \GL(n,\mathbb{F}_3) \rtimes \mathbb{F}_3^n$. \square

\end{document}